\documentclass[journal]{IEEEtran}
\IEEEoverridecommandlockouts                              
\usepackage{graphics} 
\usepackage{epsfig} 
\usepackage{mathptmx} 
\usepackage{times} 
\usepackage{amsmath} 
\usepackage{amssymb}

\usepackage{amsthm}
\usepackage{mathrsfs}
\usepackage{tikz-cd}
\usepackage{bbm}
\usepackage{mathtools}
\newtheorem{theorem}{Theorem}[section]
\newtheorem{remark}{Remark}
\newtheorem{definition}{Definition}
\newtheorem{corollary}[theorem]{Corollary}
\newtheorem{problem}{Problem}

\DeclareMathOperator*{\argmin}{arg\,min}

\allowdisplaybreaks
\title{\LARGE \bf
Cooperative Hypothesis Testing by Two Observers with \\Asymmetric Information *
}
\author{Aneesh Raghavan and John S. Baras%
\thanks{*Research supported by ARO grant W911NF-15-1-0646, by DARPA through ARO grant W911NF-14-1-0384 and by ONR grant N00014-17-1-2622.}%
\thanks{The authors are with the Department of Electrical and Computer
        Engineering and The Institute for Systems Research,
        University of Maryland, College Park, USA. Email:
        {\tt\small raghava@umd.edu, baras@umd.edu}}%
}

\begin{document}
\maketitle
\thispagestyle{empty}
\pagestyle{empty}

\begin{abstract}\label{abstarct}
 We consider the binary hypothesis testing problem with two observers. There are two possible states of nature (or hypotheses).  Observations collected by the two observers are statistically related to the true state of nature. The knowledge of joint distribution of the observations collected and the true state of nature is unknown to the observers. There are two problems to be solved by the observers: (i) true state of nature is known: find the distribution of the local information collected; (ii) true state of nature is unknown: collaboratively estimate the same using the distributions found by solving the first problem. We present four algorithms, each having two phases where the two problems are solved, with emphasis on the information exchange between the observers and resulting patterns. We prove different properties of the algorithms including the following: the probability spaces constructed as a consequence of solving the first problem are dependent on the information patterns at the observers; (ii) the rate of decay of probability of error of algorithms while solving the second problem is dependent on the information exchange between the observers. We present a numerical example demonstrating the four algorithms. 
\end{abstract}

\section{Introduction}\label{Introduction}
\subsection{Motivation}\label{Motivation}
Hypothesis testing problems arise in various areas of science, medicine, engineering, and sociology.  The standard version of the problem has been studied extensively in the literature. The inherent assumption of the standard problem is that even if there are multiple sensors collecting observations, the observations are transmitted to a single fusion center where the observations are used collectively to arrive at the belief of the true hypothesis. When multiple sensors collect observations, there could be other detection schemes as well. One possible scheme is that, the sensors could send a summary of their observations as finite valued messages to a fusion center where the final decision is made.  Such schemes are classified as ``Decentralized Detection" \cite{tsitsiklis1993decentralized}, \cite{tenney1981detection}. One of the motivations for studying decentralized detection schemes is that, when there are geographically dispersed sensors, or in environments with constrained communication (bandwidth limitations, geometric obstacles), such a scheme could lead to significant reduction in communication cost without compromising much on the detection performance. 

In probability theory, the \textit{abstract} probability space $(\Omega, \mathcal{F}, \mathbb{P})$ (a measure space) is given and always exists. The ``observables" or the random variables are measurable functions on the probability space. In a real world experiment, the measurements correspond to these observables. 
\begin{align*}
    \underbrace{(\Omega, \mathcal{F}, \mathbb{P})}_{Probability \; World} \overset{Y}{\longrightarrow} \underbrace{(\mathcal{X}, \mathcal{B}(\mathcal{X}), \mathbb{D})}_{Statistics \; World}
\end{align*}
 Let $E$ be an experiment observed by a single observer. Let the outcomes of the experiment be $O$. The observer observes a function of the outcome of the experiment, $Y = f(O)$, where $Y \in \mathcal{X}$. $(\mathcal{X}, \mathcal{T}(\mathcal{X}))$ is a topological space and $\mathcal{B}(\mathcal{X})$is the Borel $\sigma$-algebra of subsets of $\mathcal{X}$. Since $Y$ is a measurable function, $Y^{-1}(F) \in \mathcal{F}, \; \forall F \in \mathcal{B}(\mathcal{X})$. If the abstract measure, $\mathbb{P}$, is known then $\mathbb{D}(F) = \mathbb{P}(Y^{-1}(F)) \; \forall F \in \mathcal{B}(\mathcal{X})$ and is known as the law of the random variable. The probability space for the observer is $(\mathcal{X}, \mathcal{B}(\mathcal{X}), \mathbb{D})$. $\mathbb{D}$ is refereed to as a distribution when $\mathcal{X} = \mathbb{R}^{n}$ and as a measure in a general setting.
 
Problems in probability theory are often referred to as ``forward problems", i.e., given the probability space what can be said about the random variables on the space. Problems in statistics are refereed to as ``inverse problems" as often the objective is to infer the probability space given realizations of the random variables, \cite{vapnik1998statistical}.  Consider the scenario where the measure $\mathbb{P}$ is unknown. The probability space at the observer could be constructed as follows. Given the set $\mathcal{X}$, let $\mathbb{F}$ be a semiring of subsets of $\mathcal{X}$ for which the observer can assign a measure after repeatedly performing the experiment. Assuming each trial of the experiment is independent of other trials, for any $F \in \mathbb{F}$, 
\begin{align*}
    \hat{\mathbb{D}}(F) = \frac{ \# \; of \;  times \;  Y \in F }{\# \; of \; trails \; of \; experiment}
\end{align*}
 $\hat{\mathbb{D}}(\cdot)$ is a distribution on $\mathbb{F}$. Since $\hat{\mathbb{D}}(\cdot)$ is finitely additive and countably monotone, by the Caratheodory - Hahn theorem, the Caratheodory  measure $\mathbb{D}$ induced by $\hat{\mathbb{D}}$, is an extension of $\hat{\mathbb{D}}$. Let $\mathcal{B}(\mathbb{F})$ be the $\sigma$-algebra of sets which are measurable with respect to the outer measure induced by $\hat{\mathbb{D}}(\cdot)$. The probability space constructed by the observer after repeatedly observing the experiment is $(\mathcal{X}, \mathcal{B}(\mathbb{F}), \mathbb{D})$. We note that $\mathcal{B}(\mathbb{F}) = \mathcal{B}(\mathcal{X})$ if and only if $\mathcal{T}(\mathcal{X}) \subset \mathbb{F}$, i.e, the semiring $\mathbb{F}$ contains all open sets. This may not be experimentally feasible.  
 
 Suppose each trial of the experiment is observed over time and multiple observations are collected, then the observation space is $S \times T$, where $T$ denotes the instances at which the observations are collected. If $T$ is finite then the probability space construction can be done by following the methodology above. If $T$ is a countable or uncountable set, then the distributions need to satisfy the \textit{Kolmogorov Consistency} conditions \cite{liggett2010, durrett2010}. Further, the measure obtained by extending the distributions is a measure on the $\sigma$-algebra generated by the \textit{cylindrical} subsets of $S \times T$. This construction is refereed to as the \textit{Kolmogorov Construction} \cite{liggett2010, durrett2010}.

 Now we consider the scenario where the experiment is observed by two observers, Observer 1 and Observer 2.  Observer 1 observers a function of the outcome of the experiment, $Y^1 = f(O)$, while Observer 2 observes a different function $Y^2 = g(O)$ of the outcome of the experiment.  Observer 1 (Observer 2) can find the distribution of its observation $Y^1$ ($Y^2$) from the data. Neither observer can find the joint distribution of $Y^{1},Y^{2}$ as Observer 1 and Observer 2 do not know $Y^2$ and $Y^1$ respectively. Even if both of the observers share the same model for the experiment, Observer 1 (Observer 2) cannot find the distribution of $Y^{2}$ ($Y^{1}$) without knowing the $g$ ($f$) function. Hence, without sharing information, the observers cannot build the joint distribution of the observations.  To build the joint distribution, the observers could send their observations or the functions $f$ and $g$ to a central coordinator. 

In conclusion, though the abstract probability space might ``exist" and be common to the observers, the probability space at the observers are different as indicated below. When $\mathbb{P}$ is unknown, each observer could construct its own probability space by following the procedure described before. However, events which belong to $\mathcal{B}(\mathcal{X}^1 \times \mathcal{X}^2)$ are not measurable in either measure space. 
\[
  \begin{tikzcd}
     &(\mathcal{X}^1, \mathcal{B}(\mathcal{X}^1), \mathbb{D}^1)  \arrow{d}{}\\
     (\Omega, \mathcal{F}, \mathbb{P})  \arrow[swap]{ur}{Y^1} \arrow[swap]{dr}{Y^2} & ? \arrow{d}{} \arrow{u}{} \\
     & (\mathcal{X}^2, \mathcal{B}(\mathcal{X}^2), \mathbb{D}^2) \arrow{u}{}
  \end{tikzcd}
\]
The above motivates the following set of questions: (i) Is it even necessary that events in $\mathcal{B}(\mathcal{X}^1 \times \mathcal{X}^2)$ need to be measurable? (ii) What could be a suitable subset of events in $\mathcal{B}(\mathcal{X}^1 \times \mathcal{X}^2)$ that is to be measurable by each observer? (iii) What information should be exchanged by the observers to achieve the same? The questions are tied to the context in which the probability spaces are constructed. Though these problems do not naturally arise in statistics; they are inherent in multi-agent systems. This situation arises in almost all multi-agent decision making problems; though in the literature it is always assumed that $\mathbb{P}$ is known. In this paper, we study the above questions in the context of the hypothesis testing problem (as the simplest decision-making problem) and deviate from the assumption that $\mathbb{P}$ is known. In the following subsection, we present a brief survey of papers which study hypothesis testing from a multi-agent perspective. 
\subsection{Literature Survey}
In \cite{tsitsiklis1993decentralized}, the $M$-ary hypothesis testing problem is considered. A set of sensors collect observations and transmit finite valued messages to the fusion center. At the fusion center, a hypothesis testing problem is considered to arrive at the final decision. For the sensors, to decide what messages they should transmit, the Bayesian and Neyman-Pearson versions of the hypothesis testing problem are considered. The messages transmitted by the sensors are coupled though a common cost function. For both versions of the problem, it is shown that if the observations collected by different sensors conditioned on any hypothesis are independent, then the sensors should decide their messages based on the likelihood ratio test. The results are extended to the cases when the sensor configuration is a tree and when the number of sensors is large. In \cite{tenney1981detection}, the binary hypothesis testing problem is considered. The formulation considers two sensors and the joint distribution of the observations collected by the two sensors is known, to both sensors, under either hypothesis. The objective is to find an optimal decision policy for the sensors, based on the observations collected at the sensors locally, with respect to a coupled cost function. Under assumptions on the structure of the cost function, and independence of the observations conditioned on the hypothesis, it is shown that the likelihood ratio test is optimal with thresholds based on the decision rule of the alternate sensor. Conditions under which threshold computations decouple are also presented. In \cite{chamberland}, the binary decentralized detection problem over a wireless sensor network is considered. A network of wireless sensors collect measurements and send a summary individually to a fusion center. Based on the information received, the objective of the fusion center is to find the true state of nature. The objective of the study was to find the structure of an optimal sensor configuration with the formulation incorporating constraints on the capacity of the wireless channel over which the sensors are transmitting. For the scenario of detecting deterministic signals in additive Gaussian noise, it is shown that having a set of identical binary sensors is asymptotically optimal. Extensions to other observation distributions are also presented. In \cite{nayyar2011sequential}, sequential problems in decentralized detection are considered. Peripheral sensors make noisy measurements of the hypothesis and send a binary message to a fusion center. Two scenarios are considered. In the first scenario, the fusion center waits for the binary messages (i.e., the decisions) from all the peripheral sensors and then starts collecting observations. In the second scenario, the fusion center collects observations from the beginning and receives binary messages from the peripheral sensors as time progresses. In either scenario, the peripheral sensor and the fusion center need to solve a stopping time problem and declare their decision. Parametric characterization of the optimal policies are obtained and a sequential methodology for finding the optimal policies is presented. 
\subsection{Problem Description}\label{Problem Description}
We investigate the binary hypothesis testing problem from a novel and fundamental perspective. There are two possible states of nature. There are two observers, Observer 1 and Observer 2. Each observer collects its individual set of observations. The observations collected by the observers are statistically related to the true state of nature. The joint or the marginal distribution of the observations is unknown to the agents. Given the observations, the objective of the two observers is to collaboratively find the true hypothesis. The driving motivation of this paper is to understand this decentralized detection problem from scratch. More specifically the focus is to \textit{understand rigorously and at a fundamental level the construction of the underlying probability spaces, under various data exchange (communication) patterns between the two observers}. Surprisingly, this fundamental problem has not been examined carefully, and in detail, in the current literature, with the result being that several \textit{a priori} assumptions on the underlying probability spaces, widely used in the literature, are in most cases incorrect, or not justified. Further to this fundamental point, at a recent workshop in the 2023 American Control Conference \cite{varaiyaACC2023}, the surprising fact regarding the lack of such serious investigation of the underlying probability spaces and conditioning information models (i.e. $\sigma$-algebras) by each of the observers was pointed out and discussed by several authors (V. Borkar, V. Anantharam, J.S. Baras) \cite{varaiyaACC2023}. We quote from V. Anantharam's presentation ``A problem for the next generation. Develop a meaningful theory of distributed interaction when the agents have different probability assessments on the underlying sample space." Some related early papers are \cite{tenvar1984, tencast1987, tencast1988} and a more recent one is \cite{sargezyuk2019}.

Since, we do \textit{not} assume that the joint distribution of the observations is known to the observers, i.e., the observers share a common probability space,  we emphasize the \textit{details of probability space construction by each observer from the data each observer has available}. The focus is on understanding more accurately the information exchange (or not exchange) between the observers not only in order to perform collaborative detection, but even more importantly to construct the underlying probability spaces and models employed. One of the authors (Baras) has investigated the need for such detailed and careful constructions, and even the need for new probability models (i.e. von-Neumann-like and not Kolmogorov-like \cite{hintikka2002quantum}) over the years \cite{baras1979noncommutative, baras1987distributed, baras2003multiagent, baras2016multiagent}. The recent PhD thesis of one of the authors (Raghavan) \cite{raghavan2019thesis} investigated various problems closely related to the problems addressed in this paper. 

A similar problem in the study of quantum systems has gained prominence. The scenario of two independent observers conducting measurements on a joint quantum system can be modelled using two approaches: (i) a Hilbert space of tensor product form, each factor associated to one observer.  
The operators describing the observables are acting only on one tensor factor; (ii) one joint Hilbert space, requiring that all operators associated to different observers commute, i.e. are jointly measurable without causing disturbance. The problem of Tsirelson, \cite{scholz2008tsirelson}, is to decide whether all quantum correlation functions between two independent observers derived from commuting observables can also be expressed using observables defined on a Hilbert space of tensor product form. Connections between this problem and many other problems, like Connes' embedding problem \cite{junge2011connes}, Kirchberg's conjecture \cite{fritz2012tsirelson}, etc., have been established. 
\section{Problem Formulation  and Contributions}\label{Problem Formulation}
In this section, we progress towards formalizing the problem described in the previous subsection, \ref{Problem Description}, and discuss the contributions of the paper.  
\subsection{Problem Set Up}\label{Assumptions}
\begin{enumerate}
\item Both the observers operate on the same time scale; their actions are synchronized. Time is considered to be discrete and is denoted by subscripts  . Time instant $n$ is also refereed to as iteration $n$.
\item State of nature is the same for both observers. The two states of nature are represented by $0$ and $1$. The state of nature remains fixed during a given experiment but can change across experiments. The state of nature is denoted by $H$.
\item The observations collected by Observer 1 are denoted by $Y_{i}$, $Y_{i} \in S_{1}$ where $S_{1}$ is a finite set of real numbers or real vectors of finite dimension. The observations collected by Observer 2 are denoted by $Z_{i}$, $Z_{i} \in S_{2}$, where $S_{2}$ is a finite set of real numbers or real vectors of finite dimension. Let $M = |S_{1}| \times |S_{2}|$. 
\item The state of nature, $H$,  the observations, $\{Y_n, Z_n\}_{n \geq 1}$ (and functions of these observations) are considered to be random variables in an abstract probability space $(\Omega, \mathcal{F}, \mathbb{P})$, where  $\mathbb{P}$ is unknown. 
\item The knowledge of distributions of the observations (joint or marginal) under either hypothesis and the prior distribution of the hypothesis is unknown to both observers. 
\end{enumerate}
\subsection{Problems}\label{Problems}
The information collected by an observer includes observations from nature and the information communicated by the other observer. The sequence of observations collected and information received from the other observer is referred to as the information pattern at an observer. To address the last item in subsection \ref{Assumptions}, first we formulate the learning problem for the observers. 
\begin{problem}\label{Problem 1}
The Learning Problem: In this phase, the data collected by the observers comprises of realizations of (i) the true state of nature; (ii) observations which are statistically related to the true state of nature; (iii) information communicated by the other observer which is a function of the observations collected by it. For Observer 1 and 2, the data collected are realizations of the sequences of random variables
\begin{align*}
    \{H_j, Y_{j,1}, \phi^2_1(Z_{j,1}), Y_{j,2}, \phi^2_2(\{Z_{j,1},  &Z_{j,2}\}),  \\
    &\hspace{-1cm}\ldots, Y_{j,n}, \phi^2_n(\{Z_{j,l}\}^{n}_{l=1}), \ldots \}, \;  \text{and}, \\
    \{H_{j}, Z_{j,1}, \phi^1_1(Y_{j,1}), Z_{j,2}, \phi^1_2(\{Y_{j,1}, &Y_{j,2}\}),  \\
    &\ldots, Z_{j,n}, \phi^1_n(\{Y_{j,l}\}^{n}_{l=1}), \ldots\}
\end{align*} 
respectively for experiment $j$. The objective of the observers is to utilize the data over the experiments to find the local joint distribution of the random variables. 
\end{problem}
The information communicated by Observer $i$ during the learning phase is determined by the sequence of functions $\{\phi^i_n(\cdot)\}_{ n \geq 1}$. This sequence remains the same across experiments. In the testing problem, given a particular information pattern, possibly the same as the learning phase, the objective is to find the true state of nature. 
\begin{problem}\label{Problem 2}
The Testing Problem: In this phase, the data collected by the observers comprises of realizations of  (i) observations which are statistically related to the true state of nature; (ii) information communicated by the other observer which is a function of the observations collected by it and the information communicated by the former observer. For Observer 1 and 2, the data collected are realizations of the sequences of random variables
\begin{align*}
    &\{Y_1, \hat{\phi}^2_1(Z_1),\bar{\phi}^2_1(\hat{\phi}^1_1(Y_1),Z_1), Y_2, \hat{\phi}^2_2(Z_1, 
       Z_2), \bar{\phi}^2_2(\hat{\phi}^1_2(Y_1, Y_2), Z_1, \\
    &  Z_2),\ldots, Y_n, \hat{\phi}^2_n(\{Z_j\}^{n}_{j=1}), \bar{\phi}^2_n(\hat{\phi}^1_n( \{Y_j\}^{n}_{j=1}), \{Z_j\}^{n}_{j=1}, \ldots \}, \;  \text{and},  \\
    &\{Z_1, \hat{\phi}^1_1(Y_1), \bar{\phi}^1_1(Y_1,\hat{\phi}^2_1(Z_1)), Z_2, \hat{\phi}^1_2(Y_1, 
       Y_2), \bar{\phi}^1_2(Y_1, Y_2, \hat{\phi}^2_2(Z_1, \\
    & Z_2))  \ldots, Z_n, \phi^1_n(\{Y_j\}^{n}_{j=1}), \bar{\phi}^1_n( \{Y_j\}^{n}_{j=1},  \hat{\phi}^2_n(\{Z_j\}^{n}_{j=1})), \ldots\}
\end{align*}
respectively. The objective of the observers is to utilize the data to collaboratively find the true state of nature using the local knowledge, i.e., distributions obtained by solving Problem \ref{Problem 1}. 
\end{problem}
The information communicated by Observer $i$ during the testing phase is determined by $\{\hat{\phi}^i_n (\cdot)\}_{ n \geq 1}, \{\bar{\phi}^i_n (\cdot)\}_{ n \geq 1}$. It is possible that the statistical distribution knowledge to process the information received is not known to the agents, in which case the random variables are treated as \textit{exogenous} random variables, i.e, simply as real numbers without any statistical information.   
\subsection{Contributions: Algorithms}
In Section \ref{Algorithms}, we present four different algorithms to solve the problems considered. In each algorithm there are two phases: (a) Learning phase where Problem \ref{Problem 1} is solved. The true hypothesis is known, data collected is utilized  to build empirical distributions between hypothesis and the observations; (b) Testing phase where Problem \ref{Problem 2} is solved. Given a new set of observations, hypothesis testing problems are solved by the observers to find their individual beliefs about the true hypothesis. We make following observation:
\begin{itemize}
    \item The two observers possess asymmetric information and models, i.e, the distributions obtained from the learning phase are different as the information used to obtain them are different. Hence, their beliefs about the true state of nature is ``most likely " different. For them to collaborate and agree upon their beliefs, requires repeated exchange of information leading to a \textit{sequential} approach to solve the problem. 
\end{itemize}
Hence, each of the algorithms involves consensus steps for the observers to agree on their beliefs about the true hypothesis.

In Algorithm-1 (standard), subsection \ref{Centralized Approach}, the observations collected by both observers are sent to a central coordinator, the joint distribution between the observations and hypothesis is built and hypothesis testing is done using the collective set of observations. \textit{It should be noted that the joint distribution between the observations collected by the observers is found only for the purpose of comparison between the centralized and decentralized detection schemes. It is not available to observers for processing any information they receive}. 

In Algorithm-2, subsection \ref{Decentralized Approach},  each observer builds its own probability space using local observations. Hypothesis testing problems are formulated for each observer in their respective probability spaces. The observers solve the problems to arrive at their beliefs about the true hypothesis. A consensus algorithm involving exchange of beliefs is presented.

In Algorithm-3, subsection \ref{Alternative Decentralized Approach},  the observers build aggregated probability spaces by building joint distributions between their observations and the alternate observer's decisions. The decisions transmitted by the observers for probability space construction are the decisions obtained in the second approach. Hypothesis testing problems are formulated for each observer in their new probability spaces. The original decision of the observers is a function of their observations alone. The construction of the aggregated probability space enables an observer to update its information state based on the accuracy of the alternate observer. Based on the updated information state the observer updates its belief about the true hypothesis. A modified consensus algorithm is presented where the observers exchange their decision information twice; the first time they exchange their original beliefs and the second time time their updated beliefs.

In Algorithm-4, subsection \ref{Alternative Decentralized Approach with greater than 1 Bit Exchange}, we assume that the observations collected by the observers are independent conditioned on the hypothesis. In such a case the construction of the aggregated sample space is skipped. An observer receives the accuracy information (to update its information state) from the alternate observer. Hence, the observers exchange real valued information. In this approach the observers also solve the detection problem twice; once with information state obtained from the observations alone and the second time with the information state updated from the accuracy information. The consensus algorithm involves exchange of (i) original decision (ii) accuracy information (iii) updated decision.  In our previous work, \cite{raghavan2019binary}, we presented Algorithms-1, 2 and  preliminary results on the analysis of the algorithms.  

\subsection{Contributions: Analysis and Key Ideas of The Paper}\label{Key Ideas of The Paper}
In Section \ref{Analysis of the Algorithms}, we analyze properties of the distributions found and the testing algorithms prescribed in Section \ref{Algorithms}. For the learning phase of algorithms, we prove that the estimated distributions equal the true distributions. We analyze the probability space construction in the learning phase of the algorithms. We prove that the probability spaces constructed for the observers in Algorithms-2,3,4 are not \textit{similar} to the probability spaces constructed with full information exchange as in Algorithm-1.  Though the sample spaces are the same,  the measures are assigned to different $\sigma$-algebras which is dictated by the information they possess during the learning phase. Following are the key observations which are used in the proofs:
\begin{figure*}
\centering
\includegraphics[scale=0.4]{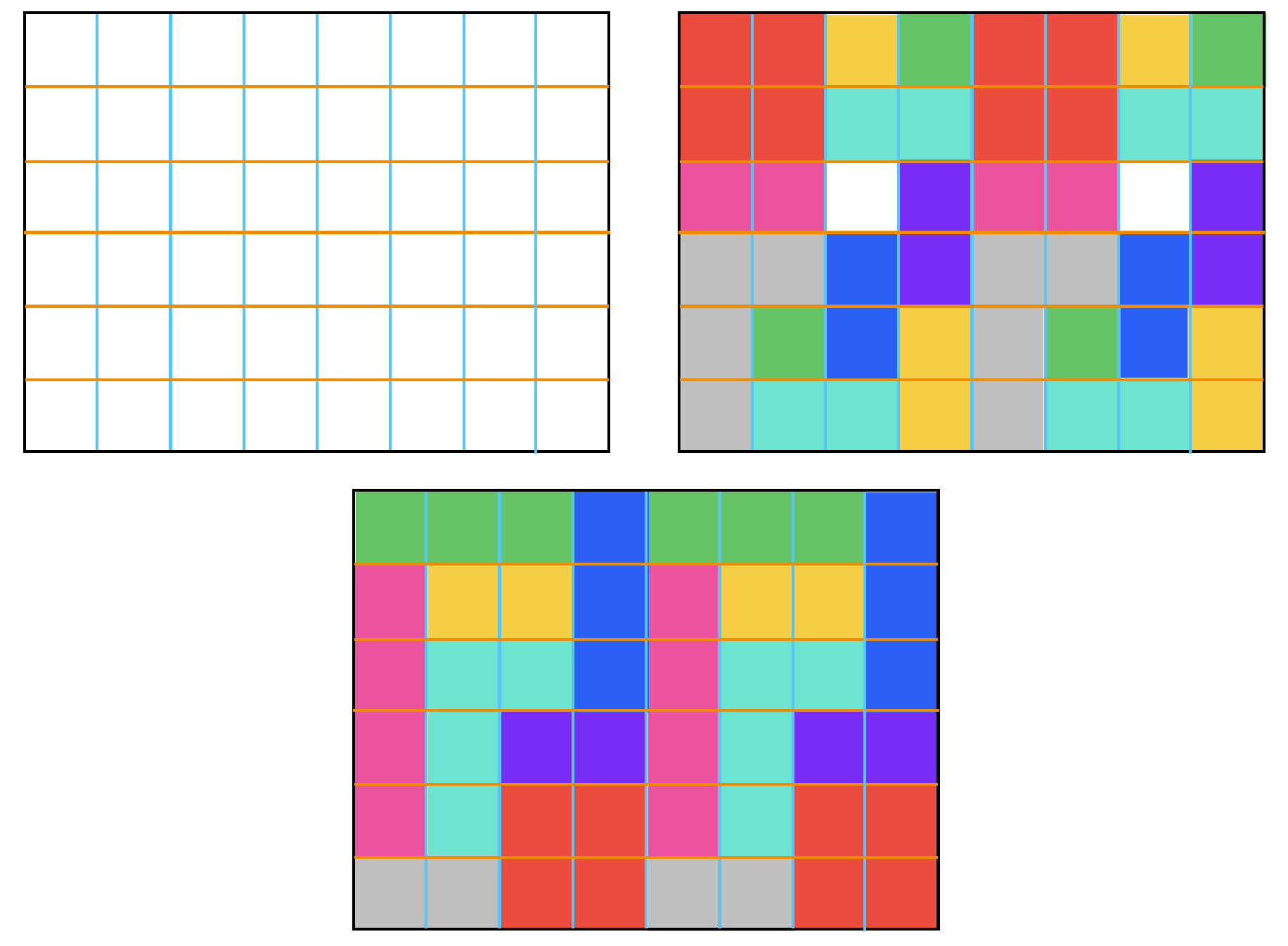}
\caption{Partition of the sample space in the learning phase for (top left) Algorithm-1 Observer 1 and 2, (top right) Algorithm-3 Observer 1 and (bottom) Algorithm-3 Observer 2.}
\label{Figure14}
\vspace{-0.7cm }
\end{figure*}
\begin{itemize}
    \item At time instant $n$, the common sample space for both agents is $\{0,1\} \times S^n_1 \times S^n_2$. Since this set is a finite set, any $\sigma$-algebra of subsets of it is obtained by union of sets in a partition of the set. A \textit{Partition} of a set, $E$, $\bar{E}$, is a finite collection of subsets of  $E$, $\bar{E} = \{E_1, \ldots E_n\}$ such that $\underset{j=1}{\overset{n}{\bigcup}} E_j = E$. If we define an \textit{Equivalence Relation} on the set $E$, the partition generated by it, $\bar{E} = \{E_1, \ldots E_n\}$, is such that $E_i \cap E_j =\varnothing$ and $\underset{j=1}{\overset{n}{\bigsqcup}} E_j = E$. This implies that, defining a measure on a $\sigma$-algebra of $\{0,1\} \times S^n_1 \times S^n_2$ generated by a partition from an equivalence relation is equivalent to defining a set function on the partition which sums to $1$, $P: \bar{E} \to [0,1] : \underset{j=1}{\overset{n}{\sum}}P( E_j) = 1$. Thus, to solve Problem \ref{Problem 1} it suffices to estimate the distributions on suitable partitions of $\{0,1\} \times S^n_1 \times S^n_2$.
    \item Different functions ($\{\phi^1_n(\cdot)\},\{\phi^2_n(\cdot)\}_{n \geq 1}$) of the observations for communication have been chosen in the four algorithms in Section \ref{Algorithms}. In Algorithm-2, subsection \ref{Decentralized Approach}, we demonstrate that for any $n$ any atomic event ( e.g. for $n=1$, $H=h, Y_1 = y_1, Z_1 = z_1$) of the sample space is not experimentally verifiable. Hence the distributions are defined on partitions whose sets contain more than $1$ element of the sample space.  In Algorithm-3, subsection \ref{Alternative Decentralized Approach}, the information exchange pattern leads to an equivalence relation on the sample space for both the observers. The corresponding partition at iteration  $n+1$, $\bar{E}_{n+1}$ (of $\{0,1\} \times S^{n+1}_1 \times S^{n+1}_2$) is a refinement of the partition at iteration $n$,  $\bar{E}_n$ (of $\{0,1\} \times S^n_1 \times S^n_2$), i.e,  set product of every set in $\bar{E}_{n}$ with  $S_1 \times S_2$ is obtained by the disjoint union of a unique collection of sets in $\bar{E}_{n+1}$. The new distribution from iteration $n$ to $n+1$  needs to be recomputed due to correlation between the information communicated in $n+1$ and all the past information. These ideas are further described in subsection \ref{Alternative Decentralized Approach Discussion}.
    \item In the Figure \ref{Figure14}, an example of the partitions generated with different information exchange has been demonstrated.  We consider $|S_1| = 6$ and $|S_2| =4$. As in Algorithm-1, if the observations collected are exchanged during the learning phase, then  every member of the sample space is a member of the partition, i.e. $\bar{E} = \{(h, y,z)\}, h \in \{0,1\}, y \in S_1, z \in S_2$ for $n=1$ (top left of the figure). Thus, there exits a unique partition which contains $2 \times 6 \times 4 = 48$ sets. However, if the information exchanged during the learning phase is only the decisions as in Algorithm-3, the partitions are not unique. One such partition is demonstrated in the figure. Further, for Observer 1, any partition has $2 \times 6 \times 2 =24$ sets while for Observer 2 any partition has  $2 \times 4 \times 2 =16$ sets for $n=1$. We note that (subsection \ref{Alternative Decentralized Approach Discussion}) the equivalence relation is such that that it does not depend on H. Hence the partitions are the same under either hypothesis. The same has been captured through the repeating color patterns. 
\end{itemize}
 In the testing phase of algorithms, we prove that consensus is achieved in Algorithms-2,3,4 for almost all sample paths. We investigate the benefits of additional data exchange in the testing phase of Algorithm-3 and establish the correlation between Algorithm-3 and Algorithm-4. The performances of these algorithms are compared by comparing the rate of decay of probability of error. We characterize the rate of decay of the probability of error in Algorithm-1 and the rate of decay of the probability of agreement on the wrong hypothesis in Algorithm-2 and prove that the former is upper bounded by the latter. We note that the study of the rate of decay of the probability of error presented in this paper is limited and is to be investigated further with advanced tools from large deviations theory, \cite{touchette2011basic}.  With respect to the computation of beliefs in the testing phase of the algorithms, we make the following remark:
 \begin{itemize}
     \item If the joint distributions of the observations under either hypothesis were to be known (i.e, the learning phase of every algorithm could  be skipped) while the information pattern in the testing phase was to be retained (Problem \ref{Problem 2}), the belief could be computed by finding the conditional probability of true state of nature given the information pattern using the joint distributions. This would require exhaustive search over the sample spaces to identify the observation sequences which lead to the information pattern, which is computationally intensive. However, in the algorithms proposed, the distributions found in the learning phase aid in finding the conditional probability using arithmetic computations, avoiding the exhaustive search. Storing the distributions, especially in Algorithm-3, could be expensive in terms of memory usage.  
 \end{itemize}
\begin{remark}
    We note that the problem considered in this paper is a decision theoretic, or a parametric hypothesis testing, analogue of the problem considered in \cite{willems2013open}. We demonstrate that the information exchanged by the observers in the learning phase dictates the probability space constructed at the observers. We prove that all subsets of the set of possible outcomes need not belong to the $\sigma$-algebra of the probability space at either observer.
\end{remark}
To summarize, the outline of the paper is as follows. In the next section, Section \ref{Algorithms},  we present the four algorithms in complete detail. In Section \ref{Analysis of the Algorithms}, we present the analysis of the algorithms including the probability space construction, rate of decay of probability of error, etc.  Simulation results are presented in section \ref{Simulation Results}.  We conclude and discuss future work in section \ref{Conclusion and Future work}. The proof of the result comparing the rate of decay of the probability of error in the testing phase of Algorithms-1,2   is presented in Section \ref{Appendix}. 
\section{Algorithms}\label{Algorithms}
In this section, we describe different algorithms to solve the problems proposed in subsection \ref{Problems}. For each algorithm, there is a learning phase where the distributions are found by solving Problem \ref{Problem 1} and subsequently suitable decision making problems are formulated. Following the learning phase, there is a testing phase where the decision making problems are solved followed by consensus steps for collaboration. 
\subsection{Algorithm-1 Centralized Approach}\label{Centralized Approach}
\subsubsection{Algorithm-1 Learning Phase}\label{Algorithm-1 Learning Phase}
\begin{figure}
\includegraphics[width=8.4cm,height=5cm]{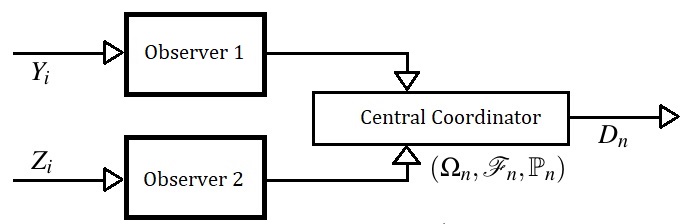}
\caption{Schematic for centralized approach} 
\label{fig:Figure1}
\end{figure}

In this approach both the observers send the data strings collected by them to a central coordinator. The central coordinator generates new strings by concatenating the observations from Observer 1, observations from Observer 2 and the true hypothesis. Thus, $\phi^1_n(\{Y_l\}^n_{l=1}) = Y_n$ and  $\phi^2_n(\{Z_l\}^n_{l=1}) = Z_n$.  From the data strings, the empirical joint distributions are found. The joint distribution when the true hypothesis is $0$ is denoted by $f_{0}(y,z)$, and when the true hypothesis is $1$ is denoted by $f_{1}(y,z)$. We assume, $0 < \mathbb{D}_{KL}(f_{0} || f_{1}) < \infty$, where $\mathbb{D}_{KL}(f_{0} || f_{1})$ denotes the Kullback-Leibler (KL) divergence between distributions $f_{0}$ and $f_{1}$ \cite{grunwald2004game, koralov2007theory}. The prior distribution of the hypothesis is denoted by $p_{h}$ for $h = 0,1$.  Let $\Omega = \{0,1\} \times S_{1} \times S_{2}$. $\omega \in \Omega$, is given by the triple $(h,y,z), h \in \{0,1\}, y \in S_{1}$ and $z \in S_{2}$. Let $\mathbb{F} = 2^{\Omega}$. Since $\Omega$ is finite it suffices to define the measure for each element in $\Omega$. Hence the measure, $\mathbb{P}^c$ is defined as follows : $\mathbb{P}^c(\omega) = p_{h}f_{h}(y,z)$.  The probability space constructed by the central coordinator is $(\Omega, \mathbb{F},\mathbb{P}^c)$.  We assume that the observations received by the observers are i.i.d conditioned on the hypothesis, in which case, the observations are random variables in the product space. The product space is defined as $(\Omega_{n}, \mathbb{F}_{n},\mathbb{P}^c_{n})$, where $\Omega_{n} = \{0,1\} \times S_{1}^{n} \times S_{2}^{n}$, $\mathbb{F}_{n} = 2^{\Omega_{n}}$ and $\mathbb{P}^c_{n}(\omega) =  p_{h}\prod^{n}_{i=1}f_{h}(y_{i},z_{i})$. The schematic for the centralized approach is shown in figure \ref{fig:Figure1}. 
Given an observation sequence $\{Y_{i},Z_{i} =y_{i},z_{i}\}^{n}_{i=1}$, the objective is to find $D_{n}:S_{1}^{n} \times S_{2}^n \longrightarrow \{0,1\}$ such that the following cost is minimized 
\begin{align*}
\mathbb{E}_{\mathbb{P}^c_{n}}[C_{10}H(1-D_{n})+C_{01}(D_{n})(1-H)],
\end{align*} 
where $H$ denotes the hypothesis random variable, $C_{10}, C_{01}$ the typical costs for decision errors. 

For the purpose of comparison of this algorithm with the remaining algorithms (e.g. subsection \ref{Alternative Decentralized Approach Discussion}), the joint probability space is extended as follows. A sample space consisting of sequences of the form $(H,(Y_{1},Z_{1}),(Y_{2},Z_{2}),(Y_{3},Z_{3}),\ldots)$ is considered. For $n \in \mathbb{N}$, let $B$ be a subset of $(\{0,1\} \times \{S_{1} \times S_{2}\}^{n})$. A cylindrical subset of $(\{0,1\} \times \{S_{1} \times S_{2}\}^{\infty})$ is:
\begin{align*}
I_{n}(B) = \{\omega \in \{0,1\} \times \{S_{1} \times & S_{2}\}^{\infty} :  \\ &(\omega(1),...,\omega(n+1)) \in B \}.
\end{align*}
Let $\mathbb{F}^{*}$ be the smallest $\sigma$-algebra generated by all cylindrical subsets of the sample space. Since the sequence of product measures $P^c_{n}$ is consistent, i.e., 
\begin{align*}
\mathbb{P}^c_{n+1}(B \times S_{1} \times S_{2}) = \mathbb{P}^c_{n}(B)\; \forall\; B \in2^{\{0,1\} \times \{S_{1}\times S_{2}\}^n},  \forall n
\end{align*} 
by the \textit{Kolmogorov extension theorem} (subsection \ref{Kolmogorov Consistency Theorem}), there exists a measure $\mathbb{P}^{c,*}$ on $(\{0,1\} \times \{S_{1}\times S_{2}\}^{\infty},\mathbb{F}^{*})$, such that, 
\begin{align*}
\mathbb{P}^{c,*}(I_{n}(B)) = \mathbb{P}_{n}(B)\;\forall\; B \in 2^{\{0,1\} \times \{S_{1}\times S_{2}\}^n},  \forall n.
\end{align*}
Hence the extended probability space at the central coordinator is given by, $(\{0,1\} \times \{S_{1} \times  S_{2}\}^{\infty},  \mathbb{F}^{*},  \mathbb{P}^{c,*})$.
\subsubsection{Algorithm-1 Testing Phase}\label{Solution Centralized Approach}
The problem formulated in the above subsection is the standard Bayesian hypothesis testing problem.  We let $\hat{\phi}^1_n(\{Y_l\}^n_{l=1}) = Y_n$, $\hat{\phi}^2_n(\{Z_l\}^n_{l=1}) = Z_n$, $\bar{\phi}^1_n(\{Y_l\}^n_{l=1}, \hat{\phi}^2_n(\{Z_l\}^n_{l=1})) = \bar{\phi}^2_n(\hat{\phi}^1_n(\{Y_l\}^n_{l=1}), \{Z_l\}^n_{l=1}) = \varnothing$. The decision policy is a threshold policy and is a function of the likelihood ratio. The likelihood ratio is defined as, $\pi_{n} = \prod^{n}_{i=1}\frac{f_{1}(y_{i}, z_{i})}{f_{0}(y_{i}, z_{i})}$. Then the decision is given by 
\[
    D_{n}=\left\{
                \begin{array}{ll}
                  1, \; \text{if},\; \pi_{n} \geq T_{c},\\
				  0, \;\text{otherwise}.
                \end{array}
              \right.
\]
where $T_{c} = \frac{C_{01}}{C_{01} + C_{10}}$. 
\subsection{Algorithm-2 Decentralized Approach} \label{Decentralized Approach}
\subsubsection{Algorithm-2 Learning Phase} \label{Algorithm-2 Learning Phase}
In this approach each observer constructs its own probability space. From the data strings collected locally, the observers find their respective empirical distributions. Thus, $\phi^1_n(\{Y_l\}^n_{l=1}) = \phi^2_n(\{Z_l\}^n_{l=1}) = \varnothing$. For Observer 1, the distribution of observations when the true hypothesis is $0$ is denoted by $f^{1}_{0}(y)$ and when the true hypothesis is $1$ is denoted by $f^{1}_{1}(y)$. Similarly, Observer 2 finds $f^{2}_{0}(z)$ and $f^{2}_{1}(z)$. We assume that the prior distribution of the hypothesis remains the same as in the previous approach. We assume, for $i=1,2$, $0 < \mathbb{D}_{KL}(f^{i}_{0} || f^{i}_{1}) < \infty$. For consistency we impose:
\begin{align*}
\sum_{z \in S_{2}} f_{h}(y,z) = f^{1}_{h}(y), \forall y \in S_{1}, h= 0,1. \\
\sum_{y \in S_{1}} f_{h}(y,z) = f^{2}_{h}(z), \forall z \in S_{2}, h= 0,1. 
\end{align*}
Based on these distributions, the probability space constructed by Observer 1 is $(\Omega^{1}, \mathbb{F}^{1},\mathbb{P}_{1})$. $\Omega^{1} = \{0,1\} \times S_{1}$, $\mathbb{F}^{1} = 2^{\Omega^{1}}$ and $\mathbb{P}_{1}(\omega) = p_{h}f^{1}_{h}(y)$. As in the previous approach, when Observer 1 receives observations which are i.i.d. conditioned on the hypothesis,  the observations are treated as random variables in the product space $(\Omega^{1}_{n}, \mathbb{F}^{1}_{n},\mathbb{P}^{1}_{n})$. For Observer 2 the probability space is $(\Omega^{2}, \mathbb{F}^{2},\mathbb{P}_{2}) = (\{0,1\} \times S_{2},2^{\Omega^{2}}, p_{h}f^{2}_{h}(z))$, while the product space is denoted $(\Omega^{2}_{n}, \mathbb{F}^{2}_{n},\mathbb{P}^{2}_{n})$. 
\begin{figure}
\includegraphics[width=8.4cm,height=5cm]{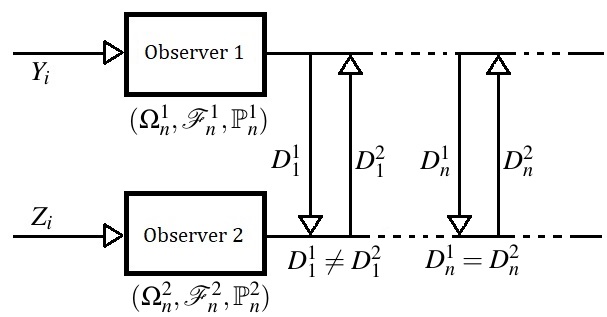}
\caption{Schematic for decentralized approach} 
\label{fig:Figure2}
\end{figure}
Given the observation sequences $\{Y_{i}=y_{i}\}^{n}_{i=1}$ and $\{Z_{i} =z_{i}\}^{n}_{i=1}$ for Observer 1 and Observer 2 respectively, the objective is to find $D^{i}_{n}:S_{i}^{n} \longrightarrow \{0,1\}$ such that following cost is minimized 
\begin{align*}
\mathbb{E}_{\mathbb{P}^{i}_{n}}[C^{i}_{10}H_{i}(1-D^{i}_{n}) + C^{i}_{01}(D^{i}_{n})(1-H_{i})],
\end{align*} 
where $H_{i}$ denotes the hypothesis random variable for observers in their respective probability spaces. 
\begin{figure}
\includegraphics[width=8.4cm,height=5cm]{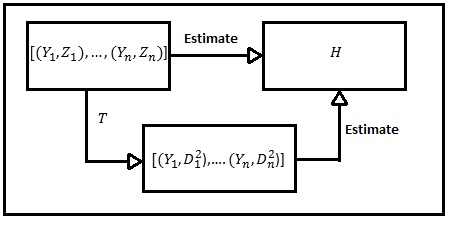}
\caption{Sufficient Statistic} 
\label{fig:Figure3}
\end{figure}

Since the decisions of the two observers need not be the same and a sequential approach with random stopping time is to proposed in the following subsection, we extend the probability spaces of the observers as follows. Since the sequences of product measures ($\{\mathbb{P}^{i}_{n}\}_{n \geq 1}$, $i=1,2$) are consistent, by the \textit{Kolmogorov extension theorem} (subsection \ref{Kolmogorov Consistency Theorem}), for $i=1,2$,  there exists measures $\mathbb{P}^{*}_{i}$ on $(\{0,1\} \times \{S_{i}\}^{\infty},\mathbb{F}^{*}_{i})$, where $\mathbb{F}^{*}_{i}$ is the $\sigma$-algebra generated by cylindrical sets in $(\{0,1\} \times \{S_{i} \}^{\infty})$, such that, 
\begin{align*}
\mathbb{P}^{*}_{i} (I^{i}_{n}(B)) = \mathbb{P}^{i}_{n}(B)\;\forall\; B \in 2^{\{0,1\} \times \{S_{i}\}^n},  \; \forall n
\end{align*}
where 
\begin{align*}
 I^{i}_{n}(B) = \{\omega \in \{0,1\} \times \{S_{i} \}^{\infty} \ni (\omega(1),...,\omega(n+1)) \in B \}.
\end{align*}
Thus, the extended probability space for Observer $i$ is $(\{0,1\} \times \{S_{i} \}^{\infty},\mathbb{F}^{*}_{i} ,\mathbb{P}^{*}_{i})$. 
\begin{remark}
Consider the scenario where $f_{h}(y,z) =f^{1}_h(y)f^{2}_h(z)$, $h=0,1$. Consider the estimation problem, where $H$ is estimated from $\{(Y_{1},Z_{1}),...,(Y_{n},Z_{n}) \}$. Let $T: S_{1}^{n} \times S_{2}^{n} \rightarrow S_{1}^{n} \times \{0,1\}^{n}$ be the mapping
\begin{align*}
    T(Y_{1},Z_{1}),...,(Y_{n},Z_{n}) = (Y_{1},D^{2}_{1}),...,(Y_{n},D^{2}_{n}).  
\end{align*}
We can consider another Bayesian estimation problem of estimating $H$ from $ (Y_{1},D^{2}_{1}),...,(Y_{n},D^{2}_{n})$.  $T$ is a sufficient statistic (Figure \ref{fig:Figure3}) for the original estimation problem if and only if
\begin{align*}
\frac{\prod^{n}_{i=1} f^{2}_{1}(z_{i})}{\sum_{z^{n}_{1} \in S_{d}}\prod^{n}_{i=1} f^{2}_{1}(z_{i})} = \frac{\prod^{n}_{i=1} f^{2}_{0}(z_{i})}{\sum_{z^{n}_{1} \in S_{d}}\prod^{n}_{i=1} f^{2}_{1}(z_{i})}, \forall \; z^{n}_{1} \in S^{2}_{d,n}, \forall \; S_{d},
\end{align*} 
where $S^{2}_{d,n}$ is a set of sequences in $S^2_{n}$, which leads to a decision sequence $\{D^{2}_{1} = d^{2}_{1},...,D^{2}_{n} = d^{2}_{n}\}$ and $z^{n}_{1}=\{z_{1}, \ldots, z_{n}\}$. The above condition is very stringent and might not be true in most cases. Even though the $T$ is not a sufficient statistic, our objective is to design a consensus algorithm based on just the exchange of decision information. The advantage of such a scheme is that, the exchange of information is restricted to $1$ bit and the observers do not have to do any other processing on their observations. 
\end{remark}
\subsubsection{Algorithm-2 Testing Phase}\label{Solution Decentralized Approach}
The \textit{information state} \cite{kumarvaraiya2015} for the observers is defined as $\psi^{i}_{n} =\mathbb{E}_{\mathbb{P}^{i}_{n}}[H| \mathcal{I}^{i}_{n}], i=1,2$, where $\mathcal{I}^{1}_{n}$ denotes the $\sigma$-algebra generated by $Y_{1},...,Y_{n}$ and $\mathcal{I}^{2}_{n}$ denotes the $\sigma$-algebra generated by $Z_{1},...,Z_{n}$. The decisions are \textit{memoryless functions} of $\psi^{i}_{n}$. More precisely, they are \textit{threshold policies}.  Let $\pi^{1}_{n} = \prod^{n}_{i=1}\frac{f^{1}_{1}(y_{i})}{f^{1}_{0}(y_{i})}$ and $\pi^{2}_{n} = \prod^{n}_{i=1}\frac{f^{2}_{1}(z_{i})}{f^{2}_{0}(z_{i})}$. Hence, $\psi^{i}_{n} = \frac{p_{1}\pi^{i}_{n}}{p_{1}\pi^{i}_{n} + p_{0}}$. For $0 < T_{i} < 1$, $\psi^{i}_{n} \geq T_{i} \Leftrightarrow \pi^{i}_{n} \geq \frac{T_{i}p_{0}}{p_{1} - T_{i}p_{1}}$. Hence the decision policy for Observer $i$ can be stated as a function of $\pi^{i}_{n}$ via: 
\[
    D^{i}_{n}=\left\{
                \begin{array}{ll}
                  1, \; \text{if},\; \pi^{i}_{n} \geq T_{i},\\
				  0, \;\text{otherwise}.
                \end{array}
              \right.
 \]
For the collaboration step,  we let 
\begin{align*}
    &\hat{\phi}^1_n(\{Y_l\}^n_{l=1}) = D^1_n, \hat{\phi}^2_n(\{Z_l\}^n_{l=1}) = D^2_n,  \\ &\bar{\phi}^1_n(\{Y_l\}^n_{l=1}, \hat{\phi}^2_n(\{Z_l\}^n_{l=1})) = \bar{\phi}^2_n(\hat{\phi}^1_n(\{Y_l\}^n_{l=1}), \{Z_l\}^n_{l=1}) = \varnothing.
\end{align*}
For an observer, a variable is said to be an exogenous random variable if it is not measurable with respect to the probability space of that observer. When Observer 1 receives the decision of Observer 2 (and vice-versa), it treats that decision as an exogenous random variable as no  statistical information is available about the new random variable. Based on this 1 bit information exchange we consider a simple consensus algorithm: Let $n =1$,
\begin{enumerate}
\item Observer 1 collects $Y_{n}$ while Observer 2 collects $Z_{n}$.
\item Based on $Y_{1},..., Y_{n}$, $D^{1}_{n}$ is computed by Observer 1, while  $D^{2}_{n}$ is computed by Observer 2 based on $Z_{1},..., Z_{n}$.
\item If $D^{1}_{n} = D^{2}_{n}$ , stop. Else increment $n$ by 1 and return to step 1. 
\end{enumerate}
The schematic of this algorithm is depicted in Figure \ref{fig:Figure2}. 
\subsection{Algorithm-3 Decentralized Approach with Two Bit Exchange}\label{Alternative Decentralized Approach}
In the previous algorithm, the decision from the alternate observer was considered as an exogenous random variable by the original observer. In this subsection, we propose a scheme where the observers build joint distributions between their own observations and the decision they receive from the alternate observer. 
\subsubsection{Algorithm-3 Learning Phase}\label{Alternative Decentralized Approach Probability Space Construction}
We let $\phi^1_n(\{Y_l\}^n_{l=1}) = D^1_n$ and $\phi^2_n(\{Z_l\}^n_{l=1}) = D^2_n$. The probability space construction for Observer 1 is described as follows: Observer 1 collects strings of finite length: $[H, Y_{1},D^{2}_{1}, Y_{2},D^{2}_{2},..., Y_{n}, D^{2}_{n}]$, where $Y_{n} \in S_{1}$ and $D^{2}_{n}$ is the decision of Observer 2, after repeating the hypothesis testing problem $n$ times. This is done by Observer 1 for every $n \in N$. $Y_{1},..., Y_{n}$ are assumed to be i.i.d. conditioned on the hypothesis and hence can be interpreted in the product space described before (section \ref{Decentralized Approach}). The decisions, $D^{2}_{1},...,D^{2}_{n}$ are obtained by Observer 2 using the decision policy described in section \ref{Solution Decentralized Approach}. Since $\pi^{i}_{n}$ are controlled Markov chains, $D^{i}_{n}$ are correlated. From the data strings, Observer 1 finds the empirical joint distribution of $\{ H, \{Y_{i}, D^{2}_{i}\}^{n}_{i=1}\}$ denoted as $\mathcal{P}_{1,n}$.  For strings of the form $(0,\{y_{i}, d^{2}_{i}\}^{i=n}_{i=1})$ or $(1,\{y_{i}, d^{2}_{i}\}^{i=n}_{i=1})$, which are not observed, measure $0$ is assigned.  Let $\mathbb{S}^{1}_{n} = \{(h,\{y_{i}, d^{2}_{i}\}^{i=n}_{i=1})\} : \mathcal{P}_{1,n}(h,\{y_{i}, d^{2}_{i}\}^{i=n}_{i=1}) >0\}$.  Hence, Observer 1 builds a family of joint distributions, $\{\mathcal{P}_{1,n}\}_{n \geq 1}$ on $2^{\mathbb{S}^{1}_{n}} \subset 2^{\{0,1\} \times \{S_{1} \times \{0,1\}\}^n}$.  We assume that the family of distributions is consistent:
\begin{align*}
\mathcal{P}_{1,n+1}(B \times S_{1} \times\{0,1\}) = \mathcal{P}_{1,n}(B)\; \forall\; B \in  2^{\mathbb{S}^{1}_{n}},  \; \forall n.
\end{align*}
Let $B$ belong to $2^{\mathbb{S}^{1}_{n}}$.  Then a cylindrical subset of $(\{0,1\} \times \{S_{1} \times \{0,1\}\}^{\infty})$ is:
\begin{align*}
I_{n}(B) = \{\omega \in \{0,1\} \times \{S_{1} \times & \{0,1\}\}^{\infty} :  \\ &(\omega(1),...,\omega(n+1)) \in B \}.
\end{align*}
Let $\mathcal{F}_{1}$ be the smallest $\sigma$-algebra such that it contains all cylindrical sets, i.e., for all $n$ and all $B$. By the \textit{Kolmogorov extension theorem} (subsection \ref{Kolmogorov Consistency Theorem}), there exists a measure $\mathcal{P}_{1}$ on $(\{0,1\} \times \{S_{1} \times \{0,1\}\}^{\infty},\mathcal{F}_{1})$ such that, 
\begin{align*}
\mathcal{P}_{1} (I_{n}(B))) = \mathcal{P}_{1,n}(B)\;\forall\; B \in 2^{\mathbb{S}^{1}_{n}} , \; \forall n,
\end{align*}
where, $I_{n}(B)$ is defined as above.
Thus, two aggregated probability spaces are constructed.  For Observer 1, \((\bar{\Omega}_{1},\mathcal{F}_{1}, \mathcal{P}_{1})\) is constructed  where $\bar{\Omega}_{1} = \{0,1\} \times \{S_{1} \times \{0,1\}\}^{\infty}$. For Observer 2, \((\bar{\Omega}_{2},\mathcal{F}_{2}, \mathcal{P}_{2})\) is constructed where $\bar{\Omega}_2 = \{0,1\} \times \{S_{2} \times \{0,1\}\}^{\infty}$. The sequence of measures $\{\mathcal{P}_{1,n}\}_{n \geq 1}$ is a function of the thresholds $T_{1}$ and $T_{2}$. Thus, when the thresholds for the decentralized scheme in \ref{Solution Decentralized Approach} change, the probability space constructed as above also changes.

Based on the new probability space constructed, the observers could find a new pair of decisions. Given the observation sequences $\{Y_{i}=y_{i}, D^{2}_{i}= d^{2}_{i}\}^{n}_{i=1}$ and $\{Z_{i} =z_{i}, D^{1}_{i} = d^{1}_{i}\}^{n}_{i=1}$ for Observer 1 and Observer 2 respectively, the objective is to find $O^{i}_{n}:\{S_{i} \times \{0,1\}\}^{n} \longrightarrow \{0,1\}$ such that the following cost is minimized 
\begin{align*}
\mathbb{E}_{\mathcal{P}_{i}}[C^{i}_{10}H_{i}(1-O^{i}_{n}) + C^{i}_{01}(O^{i}_{n})(1-H_{i})].
\end{align*} 
\subsubsection{Algorithm-3 Testing Phase}\label{Alternative Decentralized Approach Decision Scheme}
To solve the problem for Observer 1, we define a new set of information states as: 
\begin{align}
\alpha^{1}_{1} &= \mathbb{E}_{\mathcal{P}_{1}}[H_{1}|Y_{1}, D^{2}_{1}],\; \alpha^{1}_{n} = \mathbb{E}_{\mathcal{P}_{1}}[H_{1}|\{Y_{i}, D^{2}_{i}\}^{n}_{i=1}],  \nonumber \\
\alpha^{1}_{1} &= \dfrac{\mathcal{P}_{1}(D^{2}_{1} = d^{2}_{1}| Y_{1} = y_{1},H_{1} = 1)\mathcal{P}_{1}(Y_{1} = y_{1},H_{1} = 1)}{\splitdfrac{\sum_{i = 0, 1}\mathcal{P}_{1}(D^{2}_{1} = d^{2}_{1}| Y_{1} = y_{1},H_{1} = i)}{\mathcal{P}_{1}(Y_{1} = y_{1},H_{1} = i)}}\nonumber,
\end{align}
and for any $n$, 
\begin{align*}
\alpha^{1}_{n} = \hspace{-1pt} \dfrac{\splitdfrac{\mathcal{P}_{1}(Y_{n} =y_{n}, D^{2}_{n} = d^{2}_{n}| \{Y_{i} = y_{i},}{ D^{2}_{i} = d^{2}_{i}\}^{n-1}_{i=1}, H_{1} = 1)\alpha^{1}_{n-1}}}{\splitdfrac{ \sum_{j = 0, 1}\mathcal{P}_{1}(Y_{n} =y_{n}, D^{2}_{n} = d^{2}_{n}| \{Y_{i} = y_{i},}{D^{2}_{i} = d^{2}_{i}\}^{n-1}_{i=1}, H_{1} = j)[\chi_{1}(j)\alpha^{1}_{n-1} + \chi_{0}(j)(1- \alpha^{1}_{n-1})]}},
\end{align*}
where $\chi$ is the indicator function, $\chi_E(x) = 1$ if $x \in E$ and $\chi_E(x) = 0$ if $x \notin E$. 
The decision policy is :
\[
    O^{1}_{n}=\left\{
                \begin{array}{ll}
                  1, \; \text{if},\; \alpha^{1}_{n} \geq T_{3},\\
				  0, \;\text{otherwise}.
                \end{array}
              \right.
  \]
Using a similar procedure, $\{\alpha^{2}_{n}\}_{n \geq 1}$ can be defined and $\{O^{2}_{n}\}_{n \geq 1}$ can be found by Observer 2. 
\begin{figure}
\includegraphics[width=8.4cm,height=5cm]{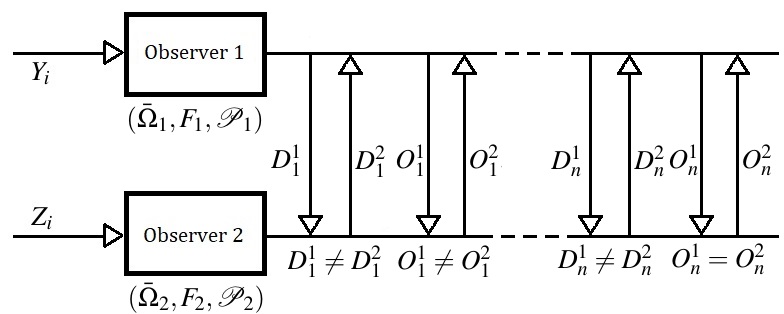}
\caption{Schematic for decentralized approach with new probability spaces} 
\label{fig:Figure5}
\end{figure}
For the collaboration step,  we let 
\begin{align*}
    &\hat{\phi}^1_n(\{Y_l\}^n_{l=1}) = D^1_n, \hat{\phi}^2_n(\{Z_l\}^n_{l=1}) = D^2_n, \bar{\phi}^1_n(\{Y_l\}^n_{l=1},  \\ &\hat{\phi}^2_n(\{Z_l\}^n_{l=1})) = O^1_n, \; \bar{\phi}^2_n(\hat{\phi}^1_n(\{Y_l\}^n_{l=1}), \{Z_l\}^n_{l=1}) =  O^2_n.
\end{align*}
The consensus algorithm can be modified from Algorithm-2 (subsection \ref{Solution Decentralized Approach}) as follows. Let $n=1$,  
\begin{enumerate}
\item Observer 1 collects $Y_{n}$ while Observer 2 collects $Z_{n}$.
\item Based on $Y_{n}, \pi^{1}_{n-1}$, $D^{1}_{n}$ is computed by Observer 1 while  $D^{2}_{n}$ is computed by Observer 2 based on $Z_{n}, \pi^{2}_{n-1}$.
\item If $D^{1}_{n} = D^{2}_{n}$ , stop. Else, $O^{1}_{n}$ is computed by Observer 1 using $\alpha^{1}_{n-1}, \{Y_{i},D^{2}_{i}\}^{n}_{i=1}$ and  $O^{2}_{n}$ is computed by Observer 2 using $\alpha^{2}_{n-1}, \{Z_{i},D^{1}_{i}\}^{n}_{i=1}$. 
\item If $O^{1}_{n} = O^{2}_{n}$, stop. Else, increment $n$ by 1 and return to step 1.
\end{enumerate}
Figure \ref{fig:Figure5} captures this approach. We note that while $\{D^{i}_n\}$ are measurable in the new probability spaces, $\{O^{i}_n\}$ is not. Hence they are treated as exogenous random variables. The interpretation of the exchange of the additional $1$ bit information is presented in subsection \ref{Algorithm 3 Testing Phase: Interpretation of Additional Data Exchange}. 
\subsection{Algorithm-4 Alternative Decentralized Approach with more than 1 Bit Exchange}\label{Alternative Decentralized Approach with greater than 1 Bit Exchange}
Motivated by the interpretation of the information states in the testing phase of Algorithm-3 (subsection \ref{Algorithm 3 Testing Phase: Interpretation of Additional Data Exchange}), we present a modification of Algorithm-3 where the construction of the ``larger" probability spaces compared to Algorithm-2 can be skipped while the essence of the improved performance of Algorithm-3 (in the testing phase) achieved through additional data exchange can still be achieved through exchange of more than $1$ bit. 
\begin{figure}
\includegraphics[width=8.4cm,height=5cm]{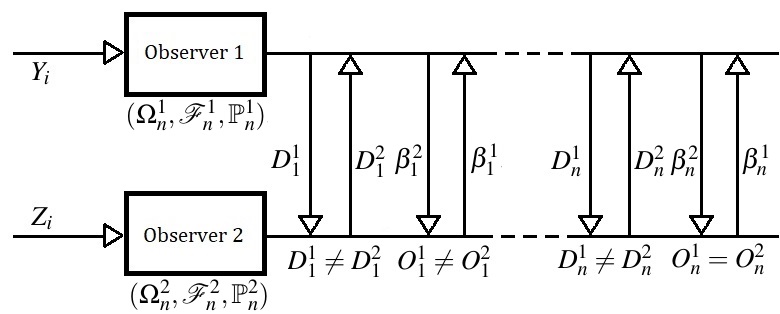}
\caption{Schematic for decentralized approach, with more than 1 bit exchange} 
\label{fig:Figure6}
\end{figure}
\subsubsection{Algorithm-4 Learning Phase}
The probability spaces constructed for the observers in learning phase of Algorithm-2, subsection \ref{Algorithm-2 Learning Phase}, are retained. Thus, the probability space for Observer $i$ is $(\{0,1\} \times \{S_{i} \}^{\infty},\mathbb{F}^{*}_{i} ,\mathbb{P}^{*}_{i})$. The decision making problems as defined in \ref{Algorithm-2 Learning Phase} and \ref{Alternative Decentralized Approach Probability Space Construction} are also retained.
\subsubsection{Algorithm-4 Testing Phase}
We define, 
\begin{align*}
    \beta^{i}_n = \frac{\mathbb{P}^{i}_{n}(D^{i}_{n} = d^{i}_{n} | \{D^{i}_{l} = d^{i}_{l}\}^{n-1}_{l=1},\;H_{i} = 0)}{\mathbb{P}^{2}_{n}(D^{i}_{n} = d^{i}_{n} | \{D^{i}_{l} = d^{i}_{l}\}^{n-1}_{l=1},\;H_{i} = 1)}, 
\end{align*}
as the estimate of accuracy of Observer $i$ by the alternate observer (subsection \ref{Algorithm 3 Testing Phase: Interpretation of Additional Data Exchange}), and, the modified information state recursively as, 
\begin{align*}
    \bar{\alpha}^{1}_1 &= \frac{\psi^{1}_{1}}{(1-\beta^{2}_{1})\psi^{1}_{1}+\beta^{2}_{1}}, \bar{\alpha}^{2}_1 = \frac{\psi^{2}_{1}}{(1-\beta^{1}_{1})\psi^{2}_{1}+\beta^{1}_{1}}, \\
    \bar{\alpha}^{1}_n &= \dfrac{\mathbb{P}_{1}(Y_{n}= y_{n}|H_{1} = 1)\bar{\alpha}^{1}_{n-1}}{\splitdfrac{\mathbb{P}_{1}(Y_{n}= y_{n}|H_{1} = 1)\bar{\alpha}^{1}_{n-1} +}{ \mathbb{P}_{1}(Y_{n}= y_{n}|H_{1} = 0)(1- \bar{\alpha}^{1}_{n-1})\beta^{2}_{n} }}, \\
     \bar{\alpha}^{2}_n &= \dfrac{\mathbb{P}_{2}(Z_{n}= z_{n}|H_{2} = 1)\bar{\alpha}^{2}_{n-1}}{\splitdfrac{\mathbb{P}_{2}(Z_{n}= z_{n}|H_{2} = 1)\bar{\alpha}^{2}_{n-1} +}{ \mathbb{P}_{2}(Z_{n}= z_{n}|H_{1} = 0)(1- \bar{\alpha}^{2}_{n-1})\beta^{1}_{n} }}, 
\end{align*}
where $\psi^{1}_{1} = \mathbb{E}_{\mathbb{P}^{*}_{1}}[H| Y_{1}]$ and $\psi^{2}_{1} = \mathbb{E}_{\mathbb{P}^{*}_{2}}[H| Z_{1}]$ (refer to subsection \ref{Solution Decentralized Approach}). For the collaboration step,  we let 
\begin{align*}
    &\hat{\phi}^1_n(\{Y_l\}^n_{l=1}) = D^1_n, \hat{\phi}^2_n(\{Z_l\}^n_{l=1}) = D^2_n, \bar{\phi}^1_n(\{Y_l\}^n_{l=1},  \\ &\hat{\phi}^2_n(\{Z_l\}^n_{l=1})) = \beta^1_n, \; \bar{\phi}^2_n(\hat{\phi}^1_n(\{Y_l\}^n_{l=1}), \{Z_l\}^n_{l=1}) =  \beta^2_n.
\end{align*}
Then the following algorithm is executed. Let $n =1$, 
\begin{enumerate}
\item Observer 1 collects $Y_{n}$ while Observer 2 collects $Z_{n}$.
\item Based on $Y_{n}$, $\pi^{1}_{n-1}$, $\pi^{1}_{n}$ is found by Observer 1. Using $\pi^{1}_{n}$, $D^{1}_{n}$ is found by Observer 1. Based on $Z_{n}$, $\pi^{2}_{n-1}$, $\pi^{2}_{n}$ is found by Observer 2. Using $\pi^{2}_{n}$, $D^{2}_{n}$ is found by Observer 2, as in subsection \ref{Solution Decentralized Approach}
\item The observers exchange their decisions. $D^{1}_{n}$ is treated as an exogenous random variable by Observer 2 while $D^{2}_{n}$ is treated as an exogenous random variable by Observer 1. If $D^{1}_{n} = D^{2}_{n} $, then stop. Else $\beta^{1}_{n}$ is sent by Observer 1 to Observer 2 while  $\beta^{2}_{n}$ is sent by Observer 2 to Observer 1. 
\item Using $Y_{n}, \bar{\alpha}^{1}_{n-1}$ and $\beta^{2}_{n}$, $\bar{\alpha}^{1}_{n}$ is computed by Observer 1 while using $Z_{n}, \bar{\alpha}^{2}_{n-1}$ and $\beta^{1}_{n}$, $\bar{\alpha}^{2}_{n}$ is computed by Observer 2. Using $\bar{\alpha}^{1}_{n}$, $O^{1}_{n}$ is computed by Observer 1 while using $\bar{\alpha}^{2}_{n}$, $O^{2}_{n}$ is computed by Observer 2 as in subsection \ref{Alternative Decentralized Approach Decision Scheme}. 
\item The observers exchange their new decisions. $O^{1}_{n}$ is treated as an exogenous random variable by Observer 2 while $O^{2}_{n}$ is treated as an exogenous random variable by Observer 1. If $O^{1}_{n} = O^{2}_{n} $, then stop. Else increment $n$ by 1 and return to step 1.  
\end{enumerate}
Figure \ref{fig:Figure6} captures the above modified algorithm. The advantage of this scheme is that the construction of the aggregated probability space is not needed. The scheme can be executed even when conditions on the joint distribution of the observations and decisions from the alternate observer do not hold, though it might not be useful. 
\section{Analysis of The Algorithms}\label{Analysis of the Algorithms}
In this section, we analyze the structural, qualitative, and, quantitative properties of the algorithms presented in the previous section. First, in subsection \ref{Convergence of Distributions}, we prove that convergence of the  distributions estimated in the learning phase to the true distributions. In subsection \ref{Alternative Decentralized Approach Discussion}, we discuss the probability space construction in the learning phase. In subsection \ref{Solution Convergence to Consensus}, we prove the convergence of the consensus step in the testing phase of the algorithms. In subsection \ref{Comparison of Error Rates}, we compare the rate of decay of the probability of error in Algorithm-1 and of the probability of agreement on the wrong belief in Algorithm-2. 
\subsection{Algorithm-1 and Algorithm-3 Learning Phase: Convergence of Distributions}\label{Convergence of Distributions}
\subsubsection{Algorithm-1}
Consider the abstract probability space $(\Omega, \mathcal{F}, \mathbb{P})$, where  $\mathbb{P}$ is unknown. Let $\{H_j, Y_{j}, Z_{j} \}_{j\in \mathbb{N}} \subset L^{\infty}(\Omega, \mathcal{F}, \mathbb{P})$ be the sequence of random variables used to estimate the distribution, $\mathbb{P}^{c}(\cdot)$, where $j$ denotes the experiment number. We assume that this sequence of random variables is i.i.d, i.e, 
\begin{align*}
    &\mathbb{P}( \{H_{l_1}, Y_{l_1}  , Z_{l_1} \}  \in E _{l_1}\cap 
    \{H_{l_2}  ,Y_{l_2}  , Z_{l_2}\} \in E_{l_2}\cap \ldots \cap  \\  
    &\{H_{l_j}  ,Y_{1_j}  , Z_{1_j} \}  \in E _{l_j}) =
    \underset{p=1}{\overset{j}{\prod}}  \mathbb{P}(\{H_{l_p} ,Y_{l_p,k}, Z_{l_p,k}\}\in E_{l_p})\\
    & \mathbb{P}(\{H_{l_1},Y_{l_1}, Z_{l_1} \}\in E ) = \mathbb{P}( \{H_{l_2},Y_{l_2} , Z_{l_2}\} \in E ),
\end{align*}
for any $j \in \mathbb{N}$, $\{l_1, \ldots, l_j\} \subset \mathbb{N}$, and, $E, E_{l_p} \subset \{0,1\} \times \{S_1 \times S_2\}$, $p=1, \ldots, j$. The estimated distributions are
\begin{align*}
    \mathbb{P}^{c}(h, y,z) = \underset{j \to \infty}{\lim} \underset{l=1}{\overset{j}{\sum}}  \dfrac{\chi_{\{h, y, z \}}(\{H_{l} ,Y_{l} , Z_{l} \})}{j}. 
\end{align*}
\begin{theorem}
    The estimated distributions equal the true distribution, i.e., 
    \begin{align*}
        \mathbb{P}^{c}(h, y,z) &= \mathbb{E}_{\mathbb{P}} [\chi_{\{h, y, z \}}(\{H_{l} ,Y_{l} , Z_{l} \})] \\
        &= \mathbb{P}[H_{l_1} =h,Y_{l_1}=y, Z_{l_1}=z]
    \end{align*}
\end{theorem}
\begin{proof}
    Follows from the \textit{strong law of large numbers}, \cite{koralov2007theory}. 
\end{proof}
The same proof can be extended to prove that the estimated measures in Algorithm-2 equal the true distributions, 
\begin{align*}
    \mathbb{P}_{1}(h, y) = \underset{j \to \infty}{\lim} \underset{l=1}{\overset{j}{\sum}}  \dfrac{\chi_{\{h, y,\}}(\{H_{l} ,Y_{l}\})}{j} = \mathbb{P}[H_{l_1} =h,Y_{l_1}=y], \\ 
    \mathbb{P}_{2}(h, z) = \underset{j \to \infty}{\lim} \underset{l=1}{\overset{j}{\sum}}  \dfrac{\chi_{\{h, z,\}}(\{H_{l} ,Z_{l}\})}{j} = \mathbb{P}[H_{l_1} =h, Z_{l_1}=z]. 
\end{align*}
\subsubsection{Algorithm-3}
Consider the abstract probability space $(\Omega, \mathcal{F}, \mathbb{P})$. Let $\{H_j, Y_{j,k}, Z_{j,k} \}_{j\in \mathbb{N}, k \in \mathbb{N}} \subset L^{\infty}(\Omega, \mathcal{F}, \mathbb{P})$ be the sequence of random variables used to estimate the distributions, $\{\mathcal{P}_{1,n}(\cdot), \mathcal{P}_{2,n}(\cdot) \}_{n \geq 1}$. Index $j$ denotes the experiment number, while index $k$ denotes the iteration; for example the random variable $Y_{j,k}$ corresponds to the  observation of Observer 1 in the $k$th iteration of experiment $j$. It is assumed that,  
\begin{align*}
    \big\{\{H_{j_1},Y_{j_1,k}, Z_{j_1,k}\}_{k \in \mathbb{N}},\{H_{j_2},Y_{j_2,k}, Z_{j_2,k}&\}_{k \in \mathbb{N}},\ldots , \\ &\{H_{j_n}, Y_{j_n,k}, Z_{j_n,k}  \}_{k \in \mathbb{N}} \big\},
\end{align*}
are independent for any $n \in \mathbb{N}$ and $\{j_1, \ldots, j_n\} \subset \mathbb{N}$, i.e, the random variables  are independent across the experiments. Thus, 
\begin{align*}
    &\mathbb{P}( \{H_{l_1}, Y_{l_1,k}  , Z_{l_1,k} \}^{n_1}_{k =1} \in E _{l_1}\cap 
    \{H_{l_2}  ,Y_{l_2,k}  , Z_{l_2,k}\}^{n_2}_{k =1} \in E _{l_2}\cap \ldots \cap  \\  
    & \{H_{l_j}  ,Y_{1_j,k}  , Z_{1_j,k} \}^{n_j}_{k =1}  \in E _{l_j}) =
    \underset{p=1}{\overset{j}{\prod}}  \mathbb{P}(\{H_{l_p} ,Y_{l_p,k}, Z_{l_p,k}\}^{n_p}_{k =1}  \in E_{l_p})
\end{align*}
 for any $j \in \mathbb{N}$, $\{l_1, \ldots, l_j\} \subset \mathbb{N}$, $\{n_1, \ldots, n_j\} \subset \mathbb{N}$, and $E_{l_p} \subset \{0,1\} \times \{S_1 \times S_2\}^{n_p}$, $p=1, \ldots, j$. The following assumption is made on the distribution. 
\begin{align*}
    \mathbb{P}(\{H_{j_1},Y_{j_1,k}, Z_{j_1,k} \}^{n}_{k =1} \in E ) = 
    \mathbb{P}( \{H_{j_2},Y_{j_2,k} , Z_{j_2,k}\}^{n}_{k =1} \in E )
\end{align*}
for any $j_1, j_2 ,n \in \mathbb{N}$, $E \subset \{0,1\} \times \{S_1 \times S_2\}^{n}$, i.e, that the random variables have the same distributions across experiments. The estimated distributions are defined as, 
 \begin{align*}
     \mathcal{P}_{1,n}(\{h, y_k, d^2_k \}^n_{k=1}) = \underset{j \to \infty}{\lim} \underset{l=1}{\overset{j}{\sum}}  \dfrac{\chi_{\{h, y_k, d^2_k \}^n_{k=1}}(\{H_{l} ,Y_{l,k} , D^2_{l,k} \}^n_{k=1})}{j},\\
    \mathcal{P}_{2,n}(\{h, z_k, d^1_k \}^n_{k=1}) = \underset{j \to \infty}{\lim} \underset{l=1}{\overset{j}{\sum}}  \dfrac{\chi_{\{h, z_k, d^1_k \}^n_{k=1}}(\{H_{l} ,Z_{l,k} , D^1_{l,k} \}^n_{k=1})}{j}.
 \end{align*}
\begin{theorem}
The estimated distributions equal the true distribution, i.e., 
\begin{align*}
     \mathcal{P}_{1,n}(\{h, y_k, d^2_k \}^n_{k=1}) = \mathbb{P}( \{H_{l}  = h ,Y_{l,k} = y_{k} , D^2_{l,k} = d^2_{k}\}^n_{k=1} ),\\
     \mathcal{P}_{2,n}(\{h, z_k, d^1_k \}^n_{k=1}) = \mathbb{P}( \{H_{l}  = h , Z_{l,k} = z_{k} , D^2_{l,k} = d^2_{k}\}^n_{k=1}),
\end{align*}
for any experiment index $l$. 
\end{theorem}
\begin{proof}
    From the definition of $\{D^1_k, D^2_k\}^n_{k=1}$, it follows that, 
 \begin{align*}
     \mathbb{P}( \{H_{l} &= h ,Y_{l,k} = y_{l,k} , D^2_{l,k} = d^2_{l,k}\}^n_{k=1} )= \\
     &\mathbb{P}(\{H_{l} = h,Y_{l,k} = y_{l,k}\}^n_{k=1}, \{Z_{l,k}\}^n_{k=1} \subset \underset{k=1}{\overset{n}{\cap}} \phi^{2^{-1}}_k(d^2_{l,k})) \\
     \mathbb{P}( \{H_{l} &= h ,Z_{l,k} = z_{l,k} , D^1_{l,k} = d^1_{l,k}\}^n_{k=1} )= \\
     &\mathbb{P}(\{H_{l} = h,Z_{l,k} = z_{l,k}\}^n_{k=1}, \{Y_{l,k}\}^n_{k=1} \subset \underset{k=1}{\overset{n}{\cap}} \phi^{1^{-1}}_k(d^1_{l,k})) ,
 \end{align*}
for any $l, n \in \mathbb{N}$. From the above and i.i.d. property of the observation sequence across experiments it follows that, for Observer 1, 
\begin{align*}
     &\mathbb{P}( \{H_{l_1}  = h_{l_1} ,Y_{l_1,k} = y_{l_1,k} , D^2_{l_1,k} = d^2_{l_1,k}\}^n_{k=1} \cap \ldots \cap \\
     &\hspace{1.2cm}\{H_{l_j}  = h_{l_j} ,Y_{l_j,k} = y_{l_j,k} , D^2_{l_j,k} = d^2_{l_j,k}\}^n_{k=1} )= \\
     &\hspace{2cm}\underset{p=1}{\overset{j}{\prod}} \mathbb{P}( \{H_{l_p}  = h_{l_p} ,Y_{l_p,k} = y_{l_p,k} , D^2_{l_p,k} = d^2_{l_p,k}\}^n_{k=1} )\\
     &\mathbb{P}( \{H_{l_1}  = h ,Y_{l_1,k} = y_{k} , D^2_{l_1,k} = d^2_{k}\}^n_{k=1})= \\
     &\hspace{2.8cm}\mathbb{P}( \{H_{l_2}  = h ,Y_{l_2,k} = y_{k} , D^2_{l_2,k} = d^2_{k}\}^n_{k=1})
\end{align*}
 Similar equations hold for Observer 2 as well. 
 \begin{align*}
     &\mathbb{P}( \{H_{l_1}  = h_{l_1} ,Z_{l_1,k} = z_{l_1,k} , D^1_{l_1,k} = d^1_{l_1,k}\}^n_{k=1} \cap \ldots \cap \\
     &\hspace{1.2cm}\{H_{l_j}  = h_{l_j} ,Z_{l_j,k} = z_{l_j,k} , D^1_{l_j,k} = d^1_{l_j,k}\}^n_{k=1} )= \\
     &\hspace{2cm}\underset{p=1}{\overset{j}{\prod}} \mathbb{P}( \{H_{l_p}  = h_{l_p} ,Z_{l_p,k} = z_{l_p,k} , D^1_{l_p,k} = d^1_{l_p,k}\}^n_{k=1} )\\
     &\mathbb{P}( \{H_{l_1}  = h ,Z_{l_1,k} = z_{k} , D^1_{l_1,k} = d^1_{k}\}^n_{k=1})= \\
     &\hspace{2.8cm}\mathbb{P}( \{H_{l_2}  = h ,Z_{l_2,k} = z_{k} , D^1_{l_2,k} = d^1_{k}\}^n_{k=1})
\end{align*}
From the strong law of large numbers, \cite{koralov2007theory}, it follows that, 
 \begin{align*}
     \mathcal{P}_{1,n}(\{h, y_k, d^2_k \}^n_{k=1}) = \mathbb{E}_{\mathbb{P}}\Big[ \chi_{\{h, y_k, d^2_k \}^n_{k=1}}(\{H_{l} ,Y_{l,k} , D^2_{l,k} \}^n_{k=1}) \Big] \\
     = \mathbb{P}( \{H_{l}  = h ,Y_{l,k} = y_{k} , D^2_{l,k} = d^2_{k}\}^n_{k=1} )\\
     \mathcal{P}_{2,n}(\{h, z_k, d^1_k \}^n_{k=1}) = \mathbb{E}_{\mathbb{P}}\Big[ \chi_{\{h, z_k, d^1_k \}^n_{k=1}}(\{H_{l} ,Z_{l,k} , D^1_{l,k} \}^n_{k=1}) \Big] \\
     =\mathbb{P}( \{H_{l}  = h , Z_{l,k} = z_{k} , D^2_{l,k} = d^2_{k}\}^n_{k=1})
\end{align*}
 \end{proof}
\subsection{Algorithm-3 Learning Phase: Probability Space Construction }\label{Alternative Decentralized Approach Discussion}
Our objective is to compare the distributions $\mathcal{P}_1, \mathcal{P}_2$ (subsection \ref{Alternative Decentralized Approach}) and $\mathbb{P}^{c,*}$ (subsection \ref{Centralized Approach}). Two distributions on the same $\sigma$-algebra can be compared; if the distributions are equal on every set of the $\sigma$-algebra then the distributions are equal. However, the considered distributions are defined on different sample spaces. Even if we find a common sample space to redefine the measures, the relationship between the corresponding $\sigma$-algebras is not clear. We investigate the same in this subsection. Comparison of distributions on two different $\sigma$-algebras is not obvious. Though two $\sigma$-algebras could have different subsets of the same sample space, if they  have the ``same " number of sets, then the distributions defined on them could be compared as below. 
\begin{definition}
Two probability spaces are said to be similar if their exists a bijection between the $\sigma$-algebras and the  measures are equal up to this bijection.  $(\Gamma_{1}, F_{1}, P_{1} ) \simeq (\Gamma_{2}, F_{2}, P_{2} )$, If  $\exists \phi: F_{1} \to F_{2 }$ such that 
\begin{align*}
& \phi(E_{1}) =\phi(\hat{E}_{1}) \Rightarrow E_{1} = \hat{E_1}\\\
&\forall E_{2} \in F_{2}, \; \exists E_{1} \in F_{1} \ni \phi^{-1}(E_{2}) =E_{1}\\
&P_{1}(E_{1}) =P_{2}(\phi(E_{1})) \; \text{and} \; P_{1}(\phi^{-1}(E_{2})) =P_{2}(E_{2})
\end{align*}
\end{definition}
Since $\phi$ also preserves the $\sigma$-algebra structure,  it can be considered to be an isomorphism between the two probability spaces.  We use the notation $\nsim$ to denote non-similarity of probability spaces.

We now present an alternate construction of the probability space for Observer 1 in Algorithm-3 so that its sample space is the same as that of the central coordinator in Algorithm-1.  Consider,  a ``modified" observation space at sample $n$,  $\{0,1\}  \times S^{n}_{1} \times S^{n}_{2}$.  Two sequences $\{h,\{y_{i},z_{i}\}^{i=n}_{i=1}\}$ and $\{h,  \{, y_{i},\bar{z}_{i}\}^{i=n}_{i=1}\}$ are said to be related, denoted by $\{h,\{y_{i},z_{i}\}^{i=n}_{i=1}\} \sim \{h,\{y_{i},\bar{z}_{i}\}^{i=n}_{i=1}\} $, if $\{z_{i}\}^{i=n}_{i=1}$ and  $\{\bar{z}_{i}\}^{i=n}_{i=1}$ lead to the same decision sequence, $\{d^{2}_{i}\}^{n}_{i=1}$.  The relation $'\sim'$ is:
\begin{itemize}
\item reflexive: $\{h,\{y_{i},z_{i}\}^{i=n}_{i=1}\} \sim \{h,\{y_{i},z_{i}\}^{i=n}_{i=1}\}$,
\item symmetric: $\{h,\{y_{i},z_{i}\}^{i=n}_{i=1}\} \sim \{h,\{y_{i},\bar{z}_{i}\}^{i=n}_{i=1}\}  \Rightarrow \{h,\{y_{i},\bar{z}_{i}\}^{i=n}_{i=1}\} \sim \{h,\{y_{i},z_{i}\}^{i=n}_{i=1}\}$,
\item transitive:$\{h,\{y_{i},z_{i}\}^{i=n}_{i=1}\}  \sim \{h,\{y_{i},\bar{z}_{i}\}^{i=n}_{i=1}\},   \{h,\{y_{i},\bar{z}_{i}\}^{i=n}_{i=1}\} \\ \sim \{h,\{y_{i},\hat{z}_{i}\}^{i=n}_{i=1}\} \Rightarrow \{h,\{y_{i},z_{i}\}^{i=n}_{i=1}\} \sim\{h,\{y_{i},\hat{z}_{i}\}^{i=n}_{i=1}\} $.
\end{itemize}
Hence $'\sim'$ is an \textit{equivalence} relation.  Let $E^{1}_{n} = \{0,1\}  \times S^{n}_{1} \times S^{n}_{2} / \sim$  be the quotient set.  Hence $E^{1}_{n}$ consists of all equivalent classes where each class contains all sequences which are equivalent to each other.  Let $\Sigma^{1}_{n}$ be the $\sigma$-algebra generated by the classes in $E^{1}_{n}$.  Since an equivalence relation partitions the set,  any pair of classes  in $E^{1}_{n}$ are mutually exclusive.  Thus,  $\Sigma^{1}_{n}$ is obtained by taking finite unions of classes in $E^{1}_{n}$.  Every sequence in $\mathbb{S}^{1}_{n}$ corresponds to a unique class in  $E^{1}_{n}$.  For every class in $E^{1}_{n}$, there exists a sequence in $\mathbb{S}^{1}_{n}$.  Hence there is a bijection,  $\phi^{1}_{n}$ from $2^{\mathbb{S}^{1}_{n}}$ to $\Sigma^{1}_{n}$.  The measure on $(E^{1}_{n}, \Sigma^{1}_{n})$ can be defined as, 
\begin{align*}
\bar{\mathcal{P}}^{1}_{n} (E) = \mathcal{P}^{1}_{n} ({\phi^{1}_{n}}^{-1}(E)),  \forall \; E\in \Sigma^{1}_{n}.
\end{align*}
Thus by construction, 
\begin{align*}
(\mathbb{S}^{1}_{n},  2^{\mathbb{S}^{1}_{n},} \mathcal{P}^{1}_{n}) \simeq (E^{1}_{n}, \Sigma^{1}_{n},\bar{\mathcal{P}}^{1}_{n} ), \; \forall n.
\end{align*}
From the consistency of $\mathcal{P}^{1}_{n}$, it follows that
\begin{align*}
\bar{\mathcal{P}}_{1,n+1}(B \times S_{1} \times S_{2}) = \bar{\mathcal{P}}_{1,n}(B)\; \forall\; B \in \Sigma^{1}_{n}, \forall n. 
\end{align*}
Let $B$ belong to $\Sigma^{1}_{n}$. Then a cylindrical subset of $(\{0,1\} \times \{S_{1} \times S_{2}\}^{\infty})$ is:
\begin{align*}
I_{n}(B) = \{\omega \in \{0,1\} \times \{S_{1} \times &S_{2}\}^{\infty} :  \\ &(\omega(1),...,\omega(n+1)) \in B \}.
\end{align*}
Let $G_{1}$ be the smallest $\sigma$-algebra such that it contains all cylindrical sets, i.e., for all $n$ and all $B$. By the \textit{Kolmogorov extension theorem} (subsection \ref{Kolmogorov Consistency Theorem}),  there exists a measure $\bar{\mathcal{P}}_{1}$ on $(\{0,1\} \times \{S_{1} \times S_{2}\}^{\infty},G_{1})$ such that, 
\begin{align*}
\bar{\mathcal{P}}_{1} (I_{n}(B)) = \mathcal{P}_{1,n}(B)\;\forall\; B \in \Sigma^{1}_{n},  \; \forall n
\end{align*}
where,  $I_{n}(B)$ is defined just above.  Define  the mapping $\phi^{1}: \mathcal{F}_{1} \to G_{1}$ as:
\begin{align*}
&\phi^{1}(I_{n}(B)) = I_{n} (\phi_{n}(B)), \; I_{n}(B) \in \mathcal{F}_{1},\\
&\phi^{1}\Big (\bigcup^{\infty}_{i=1}I_{n_i}(B_{i})\Big)= \bigcup^{\infty}_{i=1}I_{n_i}(\phi^{1}_{n_i}(B_{i})),
\end{align*}
where $I_{n}(B)$ is a cylindrical set.  It is straightforward to show that $\phi^{1}$ is a bijection by using the following properties:
\begin{align*}
(I_{n}(B))^c = I_{n}(B^c),  \; I_{n} (B_{1}) = I_{n}(B_2) \Leftrightarrow B_{1} =B_{2}.
\end{align*} 
For Observer 2,  an equivalence relation like above can be defined and $\Sigma^{2}_{n}$ can be found.  Let $G_{2}$ be the smallest $\sigma$-algebra which contains all the cylindrical sets constructed from $\{\Sigma^{2}_{n}\}^{\infty}_{n=1}$. For Observer 2, the probability space constructed is $(\{0,1\} \times \{S_{1} \times S_{2}\}^{\infty},G_{2}, \bar{\mathcal{P}}_{2})$, where $\bar{\mathcal{P}}_{2}$ is the measure obtained from \textit{Kolmogorov extension theorem} (subsection \ref{Kolmogorov Consistency Theorem}).  A bijection $\phi^2: \mathcal{F}_{2} \to G_{2}$ can also be defined.  Now let us consider the central coordinator (mentioned in section II.B).  We recall that $\mathbb{F}^{*}$ is the smallest $\sigma$-algebra which contains all the cylindrical sets constructed from $\{2^{\{0,1\} \times S^{n}_{1} \times S^{n}_{2}}\}^{\infty}_{n=1}$ and the extended probability space associated with central coordinator is $(\{0,1\} \times \{S_{1} \times S_{2}\}^{\infty},\mathbb{F}^{*},\mathbb{P}^{c,*})$. 
\begin{theorem}
Given the above constructions, 
\begin{align}
&(\bar{\Omega}_{i},\mathcal{F}_{i}, \mathcal{P}_{i}) \simeq (\{0,1\} \times \{S_{1} \times S_{2}\}^{\infty},G_{i}, \bar{\mathcal{P}}_{i})\label{Equation 1}\\
& \text{If} \; |S_{1}| > 2 \; \text{and} \; |S_{2}| >2, \; \text{then} \; ((\{0,1\} \times \{S_{1} \times S_{2}\}^{\infty},G_{i}, \bar{\mathcal{P}}_{i})\; \nsim \nonumber  \\ \hspace{-0.15cm}  &(\{0,1\} \times \{S_{1} \times S_{2}\}^{\infty},\mathbb{F}^{*},\mathbb{P}^{c,*})\label{Equation 2}
\end{align}
\end{theorem}
\begin{proof}
The first similarity in Equation (\ref{Equation 1}) follows from the constructions and the existence of the bijections $\phi^i$. 
First, we note that the sample space for the two observers and the central coordinator are the same.  \textit{The associated $\sigma$-algebras are different}.  Since the equivalence relation partitions the observation space and each equivalent class contains multiple sequences,  $\Sigma^{1}_{n}, \Sigma^{2}_{n} \subset \{2^{\{0,1\} \times S^{n}_{1} \times S^{n}_{2}}\}^{\infty}_{n=1} $.  We assume $|S_{1}| > 2 \; \text{and} \; |S_{2}| >2$ to avoid pathological  situations where the equality of the algebras could possibly hold.  For $n \in \mathbb{N}$, the number of sequences in the observation space is $2\times|S_{1}|^n \times |S_{2}|^n$.  For Observer 1, the maximum number of equivalent classes is  $2\times|S_{1}|^n \times 2^n$.  Since $2\times|S_{1}|^n \times 2^n < 2\times|S_{1}|^n \times |S_{2}|^n$,  by the \textit{Pigeon hole principle} \cite{herstein1964},  there is at least one class which contains multiple sequences from the observation space.  Let this class be denoted by $C_{n}$.  The cylindrical set corresponding to $C_n$ is
\begin{align*}
I_{n}(C_{n})= \{\omega \in \{0,1\} \times \{S_{1} \times & S_{2}\}^{\infty} :  \\ (\omega(1),...,&\omega(n+1)) \in C_{n} \},
\end{align*}
and it belongs to $G_{1}$ and $\mathbb{F}^*$.  For any $\{h,\{y_{i},z_{i}\}^{i=n}_{i=1}\} \in C_{n}$,  the cylindrical set, 
\begin{align}
&I_{n} (\{h,\{y_{i},z_{i}\}^{i=n}_{i=1}\} )= \{\omega \in \{0,1\} \times \{S_{1} \times  S_{2}\}^{\infty} :  \nonumber \\ &(\omega(1),...,\omega(n+1))= \{h,\{y_{i},z_{i}\}^{i=n}_{i=1}\}\} \label{Equation 3}
\end{align}
belongs to $\mathbb{F}^*$,  but does \textit{not} belong to $G_{1}$.  The reasoning for the same is as follows. This cylindrical set cannot be obtained from $I_{n}(C_n)$ as the set $C_{n} \backslash  \{h,\{y_{i},z_{i}\}^{i=n}_{i=1}\} \not \in \Sigma^{1}_{n}$.  We note that the partition from the equivalence relation at $n+1$ is a refinement of the partition from the equivalence relation at $n$, in the following sense,  
\begin{align*}
&\{C^{n}_{i}\},  : C^{n}_{i} \cap C^{n}_{j} = \emptyset, \bigcup_{i} C^{n}_{i} = \{0,1\}\times S^{n}_{1} \times S^{n}_{2},\\
&\{C^{n}_{i,j}\}:  C^{n}_{i} \times S_{1} \times S_{2} =  \bigcup_{j} C^{n+1}_{i,j},  C^{n+1}_{i,j} \cap C^{n+1}_{k,l} = \emptyset.
\end{align*}
Thus any cylindrical set  $I_{n+1}(B_{n+1}), B_{n+1} \in \Sigma^{1}_{n+1}$ is a subset of some cylindrical set $I_{n}(B_{n}), B_{n} \in  \Sigma^{1}_{n}$. From the above relations,  we note that the cylindrical set for a sequence like the set in Equation (\ref{Equation 3}), cannot be obtained from the union of cylindrical sets corresponding to equivalence classes at $m, \; m >n$ on the observation space,  $\{0,1\}\times S^{m}_{1} \times S^{m}_{2}$.  The finite union of cylindrical sets corresponding to equivalent classes at $m, \; m >n$,  will only result in cylindrical sets corresponding to equivalent classes at $n$. Hence, the set in Equation (\ref{Equation 3}) is neither an atomic element of the $\sigma$-algebra nor can it be obtained through the unions and intersections of other sets in the $\sigma$-algebra.  For each one of the sets,  $I_{n}(\{h,\{y_{i},z_{i}\}^{i=n}_{i=1}\}), \{h,\{y_{i},z_{i}\}^{i=n}_{i=1}\} \in C_{n}$ one cannot find a unique element to which it gets mapped in $\{I_{n}(E^{1}_{n}): E^{1}_{n} \in 2^{\Sigma^{1}_{n}})\}$.   However one can find a surjection from $\{I_{n}(E_{n}): E_{n} \in 2^{\{0,1\}\times S^{n}_{1} \times S^{n}}\}$ to $\{I_{n}(E^{1}_{n}): E^{1}_{n} \in 2^{\Sigma^{1}_{n}})\}$. Hence the set of all cylindrical subsets for Observer 1 (and Observer 2) is a strict subset of the set of all cylindrical subsets for the central coordinator,  which implies that $G_{1} \subset \mathbb{F}^{*}$ and $G_{2} \subset \mathbb{F}^{*}$, which in turn implies Equation (\ref{Equation 2}).
\end{proof}
If the aggregated probability spaces of the two observers are similar then $(\mathbb{S}^{1}_{n},2^{\mathbb{S}^{1}_{n}}, \mathcal{P}_{1,n}) \simeq (\mathbb{S}^{2}_{n},  2^{\mathbb{S}^{2}_{n},} \mathcal{P}_{2,n}), \; \forall n$.  Such a condition implies $|S_{1}|= |S_{2}|$,  and that the distribution $\mathcal{P}_{1,n}$ is a permutation of  $\mathcal{P}_{2,n}$ (and vice-versa) for all $n$.  Except for this pathological case it is safe to conclude that $((\{0,1\} \times \{S_{1} \times S_{2}\}^{\infty},G_{1}, \bar{\mathcal{P}}_{1})\; \hspace{-0.15cm} \nsim  \hspace{-0.15cm}  (\{0,1\} \times \{S_{1} \times S_{2}\}^{\infty},G_{2}, \bar{\mathcal{P}}_{2})$. Thus in the approach mentioned in section \ref{Alternative Decentralized Approach Probability Space Construction},  a probability measure is not assigned to every subset of the observation space, but is assigned to those subsets which correspond to an observable outcome.  The true $\sigma$-algebra is a coarse $\sigma$-algebra compared to the original $\sigma$-algebra.  The same concept has been emphasized in \cite{willems2013open}, i.e., models often require coarse event $\sigma$-algebras.  Through examples, it is shown that in certain experiments it might not be possible to assign measure to the \textit{Borel} $\sigma$-algebra. 
\subsection{Algorithm-2 Testing Phase: Convergence to Consensus}\label{Solution Convergence to Consensus}
\begin{theorem}
    In Algorithm-2, in the testing phase, for almost all sample paths $\omega^1 \in  \{0,1\} \times \{S_{1} \}^{\infty}$ and $\omega^2 \in  \{0,1\} \times \{S_{2} \}^{\infty}$, there exits $n(\omega^1,\omega^2) \in \mathbb{N}$ such that $D^{1}_{n(\omega^1,\omega^2) } = D^{2}_{n(\omega^1,\omega^2) }$, i.e, the observers eventually agree upon their decisions about the true state of nature. 
\end{theorem}
\begin{proof}
\(\{\psi^{i}_{n}, \mathcal{I}^{i}_{n}\}_{n \geq 1}\) are martingales in \((\{0,1\} \times \{S_{i}\}^{\infty},\mathbb{F}^{*}_{i}, \mathbb{P}^{*}_{i})\), i.e., 
\begin{align*}
    \mathbb{E}_{\mathbb{P}^{*}_{i}} [\psi^{i}_{n+1} |  \mathcal{I}^{i}_{n} ]  = \mathbb{E}_{\mathbb{P}^{*}_{i}} [\mathbb{E}_{\mathbb{P}^{*}_{i}} [H |  \mathcal{I}^{i}_{n+1} ] |  \mathcal{I}^{i}_{n} ] = \mathbb{E}_{\mathbb{P}^{*}_{i}} [ H | \mathcal{I}^{i}_{n}] = \psi^{i}_{n}.
\end{align*}
Hence by \textit{Doob}'s theorem \cite{koralov2007theory}, it follows that 
\begin{align*}
\underset{n \rightarrow \infty} \lim \psi^{i}_{n} =  H_{i}, \; \mathbb{P}^{*}_{i} \; a.s. 
\end{align*}
Hence for almost all sample paths $\omega^{i} \in \{0,1\} \times \{S_{i} \}^{\infty}$, there exist  $N(\omega^{i}) \in \mathbb{N}$ such that $D^{i}_{n} = H_{i}  \; \forall \; n \geq N(\omega^{i})$. For observer $i$, for  almost all  sample paths (or any sequence of observations), $\omega^{i}$, there exists a finite natural number $N(\omega^{i})$ such that the decision after collecting $N(\omega^{i})$ observations, or more, will always be the true hypothesis. Hence, after both observers collect $\max(N(\omega^{1}), N(\omega^{2}))$ number of samples, both their decisions will be the true hypothesis. Hence the observers achieve consensus for almost all sample paths. 
\end{proof}
Since Algorithm-3 and Algorithm-4, involve the same consensus step as Algorithm-2 (along with additional consensus steps) it follows that consensus is achieved by the observers in these algorithms as well. 
\vspace{-0.5cm}
\subsection{Algorithm 3 Testing Phase: Interpretation of Additional Data Exchange}\label{Algorithm 3 Testing Phase: Interpretation of Additional Data Exchange}
The information states used to solve the hypothesis testing problem in subsection \ref{Alternative Decentralized Approach Probability Space Construction} can be interpreted as follows. For Observer 1, $n=1$, $\alpha^{1}_{1} = \mathbb{E}_{\mathcal{P}_{1}}[H_{1}|Y_{1}, D^{2}_{1}]$ which can be expanded as follows, 
\begin{align}
\alpha^{1}_{1} &= \dfrac{\mathcal{P}_{1}(D^{2}_{1} = d^{2}_{1}| Y_{1} = y_{1},H_{1} = 1)\mathcal{P}_{1}(Y_{1} = y_{1},H_{1} = 1)}{\splitdfrac{\sum_{i = 0, 1}\mathcal{P}_{1}(D^{2}_{1} = d^{2}_{1}| Y_{1} = y_{1},H_{1} = i)}{\mathcal{P}_{1}(Y_{1} = y_{1},H_{1} = i)}}\nonumber\\
&= \frac{\psi^{1}_{1}}{(1-\beta^{2}_{1})\psi^{1}_{1}+\beta^{2}_{1}},\nonumber 
\end{align}
where, 
\begin{align*}
\beta^{2}_{1} = \frac{\mathcal{P}_{1}(D^{2}_{1} = d^{2}_{1}| Y_{1} = y_{1},H_{1} = 0)}{\mathcal{P}_{1}(D^{2}_{1} = d^{2}_{1}| Y_{1} = y_{1},H_{1} = 1)}. 
\end{align*}
The decision by Observer 1 after finding $\alpha^{1}_{1}$ is $O^{1}_{1} =1$ if $\alpha^{1}_{1} \geq T_{3} = \frac{C^{1}_{01}}{C^{1}_{01} + C^{1}_{10}}$  else $O^{1}_{1} =0$. $O^{1}_{1}$ is sent to Observer 2 which treats it as an exogenous random variable. $O^{2}_{1}$ is found by Observer 2  and sent to Observer 1 which treats it as an exogenous random variable. Suppose $\beta^{2}_{1} = 1+x$, then $\alpha^{1}_{1} = \frac{\psi^{1}_{1}}{1+ x(1-\psi^{1}_{1})}$. Consider the case where $D^{2}_{1} = 0$ and $D^{1}_{1} = 1 $. If $\beta^{2}_{1} > 1$, i.e., $x > 0$, then $\alpha^{1}_{1}  < \psi^{1}_{1}$, $\alpha^{1}_{1}$ could be less than the threshold, which implies $O^{1}_{1} = 0$. If $O^{2}_{1} = 0$ then consensus is achieved. If $\beta^{2}_{1} < 1$, i.e., $x < 0$, then $\alpha^{1}_{1}  > \pi^{1}_{1}$, $\alpha^{1}_{1}$ remains greater than the threshold, which implies $O^{1}_{1} = 1$. Hence $\beta^{2}_{1}$ could be interpreted as an estimate of the accuracy of Observer 2 by Observer 1. 

The  above expansion and interpretation of $\alpha^1_1$, motivates us to consider the expansion of the information state, $\alpha^1_n$ for all $n$ and understand that it can be defined by alternative means without construction of the extended probability space as in subsection \ref{Alternative Decentralized Approach Probability Space Construction}. Indeed, it can be achieved through the following assumption and subsequent expansion. Suppose for Observer 1 the observations collected are independent of the decisions received from Observer 2 conditioned on either hypothesis, i.e., for $j=0,1$, 
\begin{align*}
&\mathcal{P}_{1}(\{Y_{i} = y_{i}, D^{2}_{i} = d^{2}_{i}\}^{n}_{i=1} |H_{1} = j) = \\
&\mathcal{P}_{1}(\{Y_{i}= y_{i}\}^{n}_{i=1} |H_{1} = j) \mathcal{P}_{1}(\{D^{2}_{i} = d^{2}_{i}\}^{n}_{i=1} |H_{1} = j)\\
&= \left[ \prod^{n}_{i=1}\mathbb{P}_{1}(Y_{i}= y_{i}|H_{1} = j) \right] \mathcal{P}_{1}(\{D^{2}_{i} = d^{2}_{i}\}^{n}_{i=1} |H_{1} = j).
\end{align*}
Similarly for Observer 2, for $j=0,1$,
\begin{align*}
&\mathcal{P}_{2}(\{Z_{i} = z_{i}, D^{1}_{i} = d^{1}_{i}\}^{n}_{i=1} |H_{2} = j) = \\
&\left [\prod^{n}_{i=1}\mathbb{P}_{2}(Z_{i}= z_{i}|H_{2} = j)\right] \mathcal{P}_{2}(\{D^{1}_{i} = d^{1}_{i}\}^{n}_{i=1} |H_{2} = j).
\end{align*}
A sufficient condition for the above is that the observations collected by Observer 1 and Observer 2 are independent conditioned on the hypothesis. Then, $\alpha^{1}_{n}$ can be computed as follows:
\begin{align*}
\alpha^{1}_{n} &= \dfrac{\splitdfrac{\left[\prod^{n}_{i=1}\mathbb{P}_{1}(Y_{i}= y_{i}|H_{1} = 1)\right]}{\mathcal{P}_{1}(\{D^{2}_{i} = d^{2}_{i}\}^{n}_{i=1} |H_{1} = 1)p_{1}}}{\splitdfrac{\sum_{j = 0, 1}\left[ \prod^{n}_{i=1}\mathbb{P}_{1}(Y_{i}= y_{i}|H_{1} = j) \right] }{\mathcal{P}_{1}(\{D^{2}_{i} = d^{2}_{i}\}^{n}_{i=1} |H_{1} = j)p_{j}}}\\
&= \dfrac{\splitdfrac{\mathbb{P}_{1}(Y_{n}= y_{n}|H_{1} = 1) \mathcal{P}_{1}(D^{2}_{n} = d^{2}_{n} |}{ \{D^{2}_{i} = d^{2}_{i}\}^{n-1}_{i=1}, H_{1} = 1)\alpha^{1}_{n-1}}}{\splitdfrac{\sum_{j = 0, 1}\mathbb{P}_{1}(Y_{n}= y_{n}|H_{1} = j) \mathcal{P}_{1}(D^{2}_{n} = d^{2}_{n} | \{D^{2}_{i} = }{d^{2}_{i}\}^{n-1}_{i=1}, H_{1} = j)[\chi_{1}(j)\alpha^{1}_{n-1} + \chi_{0}(j)(1- \alpha^{1}_{n-1}) ]}}\\
& = \dfrac{\mathbb{P}_{1}(Y_{n}= y_{n}|H_{1} = 1)\alpha^{1}_{n-1}}{\splitdfrac{\mathbb{P}_{1}(Y_{n}= y_{n}|H_{1} = 1)\alpha^{1}_{n-1} +}{ \mathbb{P}_{1}(Y_{n}= y_{n}|H_{1} = 0)(1- \alpha^{1}_{n-1})\beta^{2}_{n} }}.
\end{align*}
In the above, the main component needed for finding $\alpha^{1}_{n}$ is, 
\begin{align*}
\beta^{2}_{n} &= \frac{\mathcal{P}_{1}(D^{2}_{n} = d^{2}_{n} | \{D^{2}_{i} = d^{2}_{i}\}^{n-1}_{i=1}, H_{1} = 0)}{\mathcal{P}_{1}(D^{2}_{n} = d^{2}_{n} | \{D^{2}_{i} = d^{2}_{i}\}^{n-1}_{i=1}, H_{1} = 1)}. \\
&= \frac{\mathbb{P}^{2}_{n}(D^{2}_{n} = d^{2}_{n} | \{D^{2}_{i} = d^{2}_{i}\}^{n-1}_{i=1},\;H_{2} = 0)}{\mathbb{P}^{2}_{n}(D^{2}_{n} = d^{2}_{n} | \{D^{2}_{i} = d^{2}_{i}\}^{n-1}_{i=1},\;H_{2} = 1)}
\end{align*}
$\beta^{2}_{n}$ can be computed by Observer 2 as above using the distributions estimated in the learning phase of Algorithm-2 and the following expansion. Given $\{ d^{2}_{i}\}^{n}_{i = 1}$,
\begin{align*}
& \mathbb{P}^{2}_{n} (\{D^{2}_{i} = d^{2}_{i}\}^{n}_{i = 1}) = \underset{\{j=0,1\}}\sum \underset{\{z_{1} \in S_{2} : \chi_{1}(d^{2}_{1})(\pi^2_1 \geq T_{2}) + \chi_{0}(d^{2}_{1})(\pi^2_1 < T_{2})\}} \sum  \\ & \underset{\{z_{2} \in S_{2} : \chi_{1}(d^{2}_{2})(\pi^2_2 \geq T_{2}) + \chi_{0}(d^{2}_{2})(\pi^2_2 < T_{2})\}} \sum  \ldots 
\\ &\underset{\{z_{n} \in S_{2} : \chi_{1}(d^{2}_{n})(\pi^2_n \geq T_{2}) + \chi_{0}(d^{2}_{n})(\pi^2_n < T_{2})\}} \sum \left[\prod^{n}_{i=1}\mathbb{P}_{2}(Z_{i} = z_{i}|H_{2}=j)\right]p_{j}.
\end{align*}
$\beta^{2}_{n}$ can be interpreted as the accuracy of Observer 2. When the decision is equal to the true hypothesis with high probability,  $\beta^{2}_{n}>> 1$ when ($H=0$) and $\beta^{2}_{n}<< 1$ when ($H=1$). $\alpha^1_{n-1}$ is the belief of Observer 1 that the true hypothesis is $1$. When $\beta^{2}_{n}<< 1$, $\alpha^1_{n}$ gets closer to $1$ as $\alpha^1_{n}\thickapprox \dfrac{\mathbb{P}_{1}(Y_{n}= y_{n}|H_{1} = 1)\alpha^{1}_{n-1}}{\mathbb{P}_{1}(Y_{n}= y_{n}|H_{1} = 1)\alpha^{1}_{n-1}}$. Similarly, when $H=0$ and $\beta^{2}_{n}>> 1$, $\alpha^1_{n}\thickapprox \dfrac{1}{\mathbb{P}_{1}(Y_{n}= y_{n}|H_{1} = 0)(1- \alpha^{1}_{n-1})\beta^{2}_{n}}$ which is closer to $0$. Thus, the accuracy of the alternate observer aids the information state of the former observer towards the correct hypothesis. This interpretation of the information state of Algorithm-3 testing phase leads to Algorithm-4. 
\subsubsection{Remark on Memory Vs Computation in Algorithm-3 and Algorithm-4}
In Algorithm-3, computing and storing the collection of distributions, $\{\mathcal{P}_{i,n}\}_{n \geq 1}, \; i =1,2$ might be expensive. However, computing the information states, $\alpha^{i}_n$, involves only arithmetic computations, specifically addition, inversion, and multiplication. In Algorithm-2 (and Algorithm-4) finding the distributions and storing them is less expensive compared to Algorithm-3. However in Algorithm-4, computing $\beta^{i}_n$, is computationally expensive as it requires exhaustive search over $S^n_1$ or $S^n_2$ to find appropriate observation sequences which lead to the given decision sequence. Hence, Algorithm-3 is memory intensive while Algorithm-4 is computation intensive. 
\subsection{Algorithm-1 Vs Algorithm-2: Comparison of Error Rates}\label{Comparison of Error Rates}
In this section, we study the rate at which probability of error decays as more observations are collected. We compare the rates achieved using Algorithm-1 and Algorithm-2. 
\subsubsection{Algorithm-1}\label{Comparison of Error Rates Centralized Approach}
In this subsection we define the probability of error and its optimal rate of decay for the centralized approach. Let,
\begin{align*}
&\mathcal{A}_{n} = \{(Y_{i},Z_{i})^{n}_{i=1} \in S^{n}_{1} \times S^{n}_{2} \ni D_{n} =1 \},\\
&\kappa_{n}  = \mathbb{P}_{n}(\mathcal{A}_{n} | H = 0), \xi_{n}  = \mathbb{P}_{n}(\mathcal{A}^{c}_{n} | H = 1),
\end{align*}
where $\mathcal{A}^{c}_{n} $ is the complement of $\mathcal{A}_{n} $. Then, the probability of error $\gamma_{n}$ is 
\begin{align*}
\gamma_{n} = \mathbb{P}_{n}(D_{n} \neq H) = p_{0}\kappa_{n} + p_{1}\xi_{n}. 
\end{align*}
The optimal rate of decay of the probability of error for the centralized approach is defined as, 
\begin{align*}
R^{*}_{c} = \underset{n \rightarrow \infty} \lim -\frac{1}{n} \log_{2}\left ( \underset{\mathcal{A}_{n} \subseteq S^{n}_{1} \times S^{n}_{2} }\min\gamma_{n} \right )
\end{align*}
We define the following distributions which will help us characterize $R^{*}_{c}$.  For $h=0,1, \tau_{h} \geq 0$ define  
\begin{align}
\mathbb{Q}^{h}_{\tau_{h}}(y,z) = \frac{(f_{h}(y,z))^{1-\tau_{h}}(f_{1-h}(y,z))^{\tau_{h}}}{\sum_{y,z}(f_{h}(y,z))^{1-\tau_{h}}(f_{1-h}(y,z))^{\tau_{h}}}.\label{Equation 4}
\end{align}
Then, 
\begin{align}
R^{*}_{c}= \underset{\tau_{0},\tau_{1} \geq 0} \max \;\; \min \left[\mathbb{D}_{KL}(\mathbb{Q}^{0}_{\tau_{0}} || f_{0}), \mathbb{D}_{KL}(\mathbb{Q}^{1}_{\tau_{1}} || f_{1})\right]. \label{Equation 5}
\end{align}
\subsubsection{Algorithm-2}\label{Comparison of Error Rates Decentralized Approach}
To compare the rate of decay of the probability of error in the second approach to that in the first approach, we consider that in the second approach there is a hypothetical central coordinator where the joint distribution was built. Let,
\begin{align}
&\mathcal{B}^{1}_{n} = \{(Y_{i},Z_{i})^{n}_{i=1} \in S^{n}_{1} \times S^{n}_{2} \ni D^{1}_{n} =1 \; \text{and} \; D^{2}_{n} =1 \}.\label{Equation 6}\\
&\mathcal{B}^{2}_{n} = \{(Y_{i},Z_{i})^{n}_{i=1} \in S^{n}_{1} \times S^{n}_{2} \ni D^{1}_{n} =0 \; \text{and} \; D^{2}_{n} =0 \}.\label{Equation 7}\\
&\mu_{n}  = \mathbb{P}_{n}(\mathcal{B}^{1}_{n} | H = 0), \nu_{n}  = \mathbb{P}_{n}(\mathcal{B}^{2}_{n} | H = 1). \nonumber
\end{align}
For the probability space $(\Omega_{n}, \mathbb{F}_{n},\mathbb{P}_{n})$, the algebra $\mathbb{F}_{n}$ contains all possible subsets of the product space. Hence $\mathcal{B}^{1}_{n}$ and $\mathcal{B}^{2}_{n}$ are measurable sets. Note that, the decision regions $\mathcal{B}^{1}_{n}$ and $\mathcal{B}^{2}_{n}$ depend on thresholds $T_{1}$ and $T_{2}$ respectively; by changing the thresholds different decision regions can be generated. Given a fixed number of samples, $n$, to both the observers, let $D^{1}_{n}$ and $D^{2}_{n}$ denote their decisions. The probability that the two observers agree on the wrong belief is, $\rho_{n}$,  
\begin{align*}
\rho_{n} = \mathbb{P}_{n}(D_{c} \neq H) = p_{0}\mu_{n} + p_{1}\nu_{n}, 
\end{align*}
where $D_{c} = D^{1}_{n} = D^{2}_{n}$. The rate of decay of the probability of agreement on the wrong belief for the decentralized approach is defined as:
\begin{align*}
R_{d} = \underset{n \rightarrow \infty} \lim -\frac{1}{n} \log_{2} \left(\rho_{n}\right).
\end{align*}
The optimal rate of decay of the probability of agreement on the wrong belief for the decentralized approach is defined by optimizing over thresholds :
\begin{align*}
R^{*}_{d} = \underset{n \rightarrow \infty} \lim -\frac{1}{n} \log_{2} \left( \underset{\mathcal{B}^{1}_{n}, \mathcal{B}^{2}_{n} \subseteq S^{n}_{1} \times S^{n}_{2} }\min\rho_{n}\right).
\end{align*}
Define, the following probability distributions: for $h=0,1$,  $\lambda_{h} \geq 0, \sigma_{0}\geq 0$
\begin{align}
&\mathbb{Q}^{h}_{\lambda_{h},\sigma_{h}}(y,z) = \frac{K_{h}f_{h}(y,z)(f^{1}_{0}(y))^{s(h)\lambda_{h}}(f^{2}_{0}(z))^{s(h)\sigma_{h}}}{(f^{1}_{1}(y))^{s(h)\lambda_{h}}(f^{2}_{1}(z))^{s(h)\sigma_{h}}}, \nonumber \\
& K_{h} = \left[\sum_{y,z}\frac{f_{h}(y,z)(f^{1}_{0}(y))^{s(h)\lambda_{h}}(f^{2}_{0}(z))^{s(h)\sigma_{h}}}{(f^{1}_{1}(y))^{s(h)\lambda_{h}}(f^{2}_{1}(z))^{s(h)\sigma_{h}}}\right]^{-1} , \label{Equation 8}
\end{align}
where $s(h) =1$ if $h=1$ and $s(h )=-1$ if $h=0$.  
Then,
\begin{align}
R^{*}_{d} = \underset{\lambda_{h} \geq 0, \sigma_{h} \geq 0, h=0,1} \max  \min \left[\mathbb{D}_{KL}(\mathbb{Q}^{0}_{\lambda_{0}, \sigma_{0}} || f_{0}), \mathbb{D}_{KL}(\mathbb{Q}^{1}_{\lambda_{1}, \sigma_{1}} || f_{1})\right]. \label{Equation 9}
\end{align}
\begin{theorem}
Given the above definitions, and if $f_{0}(y,z) = f^{1}_{0}(y)f^{2}_{0}(z)$ and $f_{1}(y,z) = f^{1}_{1}(y)f^{2}_{1}(z)$, then 
\begin{align}
R^{*}_{d}  \geq R^{*}_{c}.\label{Equation 10}
\end{align} 
\end{theorem}
For the proof of Equations (\ref{Equation 4}), (\ref{Equation 5}), (\ref{Equation 8}), (\ref{Equation 9}), and the above result we refer to the Appendix. 
\begin{corollary}
If the inequality is strict in  Equation (\ref{Equation 10}),  then $\exists \quad N$ such that 
\begin{align*}
\underset{\mathcal{B}^{1}_{n}, \mathcal{B}^{2}_{n} \subseteq S^{n}_{1} \times S^{n}_{2} }\min\rho_{n}  < \underset{\mathcal{A}_{n} \subseteq S^{n}_{1} \times S^{n}_{2} }\min\gamma_{n},  \; \forall n \geq N.
\end{align*}
\end{corollary}
\subsubsection{Probability of Error}\label{Comparison of Error Rates Probability of Error}
First, we note that the number of samples collected by the two observers before they stop is random. Let the random number of samples collected by the observers before they stop be $\tau_{d}$. $\tau_{d}$ is a stopping time of the filtration generated by the sequence, $\{Y_{n},Z_{n}\}_{n \in \mathbb{N}}$, and hence it is a random variable in the extended joint probability space,$(\{0,1\} \times \{S_{1} \times S_{2}\}^{\infty},\mathbb{F}^{*},\mathbb{P}^{c,*}$). Let $D_{\tau_{d}}$ denote the decision at consensus. We note that $D_{\tau_{d}}$ is also a random variable in the extended joint probability space. Then the probability of error for the consensus scheme is:
\begin{align*}
&\mathbb{P}^{c,*}(D_{\tau_{d}} \neq H)= \sum^{\infty}_{n=1}\mathbb{P}^{c,*}((D_{\tau_{d}} \neq H) \cap  \tau_{d} =n)\\
&=\sum^{\infty}_{n=1}\mathbb{P}^{c,*}((\{D^{1}_{i} \neq D^{2}_{i}\}^{n-1}_{i=1}) \cap (D^{1}_{n} = D^{2}_{n})\cap (D^{1}_{n} \neq H))\\
&=\sum^{\infty}_{n=1}\mathbb{P}_{n}((\{D^{1}_{i} \neq D^{2}_{i}\}^{n-1}_{i=1}) \cap (D^{1}_{n} = D^{2}_{n})\cap (D^{1}_{n} \neq H))\\
&\leq \sum^{\infty}_{n=1}\mathbb{P}_{n}((D^{1}_{n} = D^{2}_{n}) \cap ( D^{1}_{n} \neq H))  \approx \sum^{\infty}_{n=1} 2^{-nR_{d}} = \frac{1}{2^{R_{d}}-1}.
\end{align*}
The first equality follows from the law of total probability. The second equality follows from the stopping rule of the consensus algorithm. Let $B = \{\{h\}\times(y_{i},z_{i})^{n}_{i=1} \in \{0,1\} \times S^{n}_{1} \times S^{n}_{2} \ni \{d^{1}_{i} \neq d^{2}_{i}\}^{n-1}_{i=1}, d^{1}_{n} = d^{2}_{n} \neq h\}$. $\omega$ is such that $\{D^{1}_{i}(\omega) \neq D^{2}_{i}(\omega)\}^{n-1}_{i=1}, D^{1}_{n}(\omega) = D^{2}_{n}(\omega)\neq H\}$ is the set of sequences for which $\{(H,(Y_{i},Z_{i})^{n}_{i=1})\} \in B$,  which corresponds to cylindrical set with, $B$, $B \in \{0,1\} \times S^{n}_{1} \times S^{n}_{2}$. Hence the third equality follows. The usefulness of the approximate upper bound for the probability of error depends on $R_{d}$. By choosing different values for the thresholds, $T_{1}$ and $T_{2}$, different values of $R_{d}$ can be obtained. Hence the upper bound is a function of the thresholds. Given the distributions under either hypotheses and the thresholds for the observers, it is difficult to numerically compute the probability of error (given by the first equality above) as it requires an exhaustive search over the observation space for high values of $n$. We estimate the probability of error empirically using simulations and the results are presented in section \ref{Simulation Results}.  

\begin{figure}
\includegraphics[width=8.4cm,height=5cm]{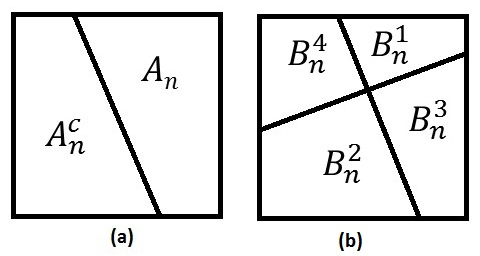}
\caption{Observation space divided in to (a) two regions (b)  four regions} 
\label{fig:Figure4}
\end{figure}

The result of Equation (\ref{Equation 10}) can be interpreted as follows: Given a fixed number of samples $n$, the minimum probability of error achieved in the centralized approach is approximately $2^{-nR^{*}_{c}}$. Given the same number of samples for the decentralized approach, the minimum probability that the observers agree and are wrong is  $2^{-nR^{*}_{d}}$. Hence the above result implies that, for sufficiently large $n$, the minimum probability of the observers agreeing and being wrong in the decentralized approach is upper bounded by the minimum probability of error in the centralized approach. The result can be understood heuristically as follows: The observation space after collecting $n$ observations is $Y^{n} \times Z^{n}$. In the centralized approach, the observation space is divided into two regions, one where the decision is 1 ($A_{n}$), and the other is where the decision is 0 ($A^{c}_{n}$), (Figure \ref{fig:Figure4}a). In the decentralized approach, the observation space is divided into four regions (Figure \ref{fig:Figure4}b) : (1) Decision of Observer 1 is 1 and Decision of Observer 2 is 1 ($B^{1}_{n}$); (2) Decision of observer 1 is 0 and Decision of observer 2 is 0   ($B^{2}_{n}$); (3) Decision of observer 1 is 0 and Decision of observer 2 is 1 ($B^{3}_{n}$); (4) Decision of Observer 1 is 1 and Decision of observer 2 is 0 ($B^{4}_{n}$). The observers can be wrong only in regions $B^{1}_{n}$ and $B^{2}_{n}$ depending on the true hypothesis. Since the measure of regions $B^{1}_{n}$ and $B^{2}_{n}$ are likely going to be less than the measure of the regions $A_{n}$ or $A^{c}_{n}$ the probability of the observers agreeing and being wrong in the second approach is going to be likely less than the probability of error of the central coordinator. 
\begin{remark}
The consensus algorithm presented in section \ref{Solution Decentralized Approach} translates to considering sets of the form $\{(Y_{i},Z_{i})^{n}_{i=1} \in S^{n}_{1} \times S^{n}_{2} \ni \{D^{1}_{i} \neq D^{2}_{i}\}^{n-1}_{i=1}, D^{1}_{n} = D^{2}_{n}=1 \}$ and  $\{(Y_{i},Z_{i})^{n}_{i=1} \in S^{n}_{1} \times S^{n}_{2} \ni \{D^{1}_{i} \neq D^{2}_{i}\}^{n-1}_{i=1}, D^{1}_{n} = D^{2}_{n}=0 \}$ in section \ref{Comparison of Error Rates Decentralized Approach}.  It is essential that these sets can be equivalently captured by a set of distributions in the probability simplex in $\mathbb{R}^{|S_{1} \times S_{2}|}$ for computation of the rates as done in section \ref{Appendix}.  Since these sets cannot be equivalently captured by a set of distributions, we consider a superset of the sets described in (\ref{Equation 6}) and (\ref{Equation 7}). Thus, we are able to only obtain an upper bound for the probability of error in section \ref{Comparison of Error Rates Probability of Error}. 
\end{remark}
\begin{remark}
There could be other possible schemes for decentralized detection. For example each observer could individually solve a stopping time problem. The times at which they stop are functions of the probability of error they want to achieve. Hence the observers stop at random times and send their decision information when they stop. The same consensus protocol could be used, i.e., the observers stop only when they both arrive at the same decision. In this scheme the probability of error of the decentralized scheme is upper bounded by the max of the probability of error of the individual observers.  
\end{remark}
\subsubsection{Remark on the Rate of Convergence of Algorithm-3}
Even though the two observers do not share a common probability space, to compare the probability error we consider the same joint distribution as in Algorithm-1. The probability of error is given by:
\begin{align*}
\mathbb{P}_{e,n} = & \underset{\{y^{n},z^{n} \ni (\alpha^{1}_{n} \geq T_{3} \; \cap \; \alpha^{2}_{n} \geq T_{4} ) \}} \sum f_{0}(y,z) + \\ & \underset{\{y^{n},z^{n} \ni (\alpha^{1}_{n} < T_{3} \; \cap \; \alpha^{2}_{n} < T_{4}) \} } \sum f_{1}(y,z), 
\end{align*}
where $T_{4} = \frac{C^{2}_{01}}{C^{2}_{10}+C^{2}_{01}}$. In this scenario, it is difficult to characterize the error rate. In the previous subsection, for Algorithm-2, the method of types was used to find the error rate. The sets used to characterize the error rate would now depend on the decision sequence from the alternate observer. For a particular type, there could be multiple decision sequences. Hence, the same approach cannot be extended. The advantage of this algorithm is that it has faster rate of convergence due to step 4 of the consensus algorithm. However, the characterization of the rate of decay in Algorithm-3 and Algorithm-4  remains an open problem. 
\section{Simulation Results}\label{Simulation Results}
\begin{figure}
\includegraphics[width=8.4cm,height=5cm]{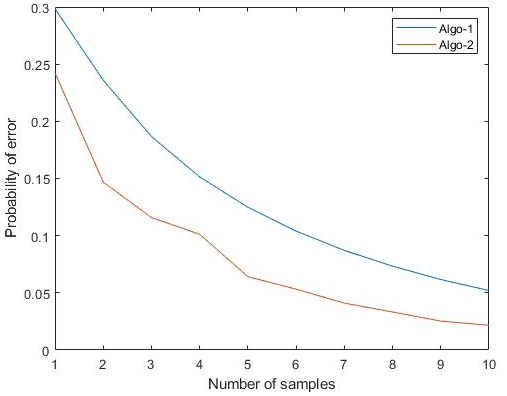}
\caption{Probability of error / conditional probability of agreement on wrong belief )  vs number of samples} 
\label{fig:Figure7}
\end{figure}

\begin{figure}
\includegraphics[width=8.4cm,height=5cm]{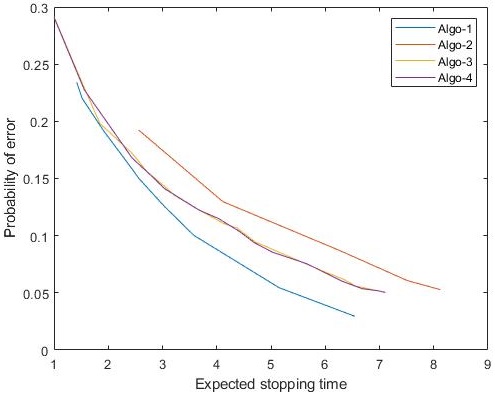}
\caption{Probability of error vs expected stopping time} 
\label{fig:Figure8}
\end{figure}

Simulations were performed to evaluate the performance of the algorithms. The setting is described as follows. The cardinality of the sets of observations collected by Observers 1 and 2 are $3$ and $4$ respectively. The joint distribution of the observations under either hypothesis is given in Table 1. Note that under either hypothesis, the observations received by the two observers are independent. 
\begin{table}[h!]
\begin{center}
\begin{tabular}{||c || c|| c ||c ||c ||} 
 \hline
 $f_{0}(y,z)$  & $Z=1$ & $Z=2$ &$Z =3$& $Z=4$ \\
 \hline
 $Y=1$ & $0.02$ &$0.05$ & $0.07$ &$0.06$\\
 \hline
 $Y=2$ & $0.03$ &$0.075$& $0.105$ &$0.09$ \\
 \hline
 $Y=3$ & $0.05$ &$0.125$ &$0.175$ &$0.15$ \\
 \hline
\end{tabular}
\begin{tabular}{||c || c|| c ||c ||c ||} 
 \hline
 $f_{1}(y,z)$  & $Z=1$ & $Z=2$ &$Z =3$& $Z=4$ \\
 \hline
 $Y=1$ & $0.18$ &$0.135$ & $0.09$ &$0.045$\\
 \hline
 $Y=2$ & $0.1$ &$0.075$& $0.05$ &$0.025$ \\
 \hline
 $Y=3$ & $0.12	$ &$0.09$ &$0.06$ &$0.03$ \\
 \hline
\end{tabular}
\caption{Table 1. Joint distribution of observations under either hypothesis }
\label{table:1}
\end{center}
\vspace{-1cm}
\end{table} 
The prior distribution of the hypothesis was considered to be $p_{0} =0.4$ and $p_{1} =0.6$. $\mathbb{D}_{KL}(f_{1}||f_{0}) = 0.7986 $ and $\mathbb{D}_{KL}(f_{0}||f_{1}) = 0.7057 $. The empirical probability of error achieved by using the centralized scheme as $n$ increases has been plotted in Figure \ref{fig:Figure7} (Algo-1). The empirical probability of the observers agreeing on the wrong belief conditioned on the observers agreeing in the decentralized scheme (\ref{Solution Decentralized Approach}) has been plotted in Figure \ref{fig:Figure7} (Algo-2). In order to construct the aggregated sample space, the joint distribution of the observations and decision was found by the frequentist approach. $2 \times 10^7$ samples were used to construct the aggregated sample space. The empirical probability of error achieved by the  centralized sequential hypothesis testing scheme (using sequential probability ratio test), by the decentralized scheme in section \ref{Solution Decentralized Approach}, by the decentralized scheme in section \ref{Alternative Decentralized Approach Decision Scheme}, by the decentralized scheme in section \ref{Alternative Decentralized Approach with greater than 1 Bit Exchange}, has been plotted against the expected stopping time in Figure \ref{fig:Figure8}, Algo-1, Algo-2, Algo-3, and Algo-4  respectively. It is clear that the centralized sequential scheme performs the best among the four schemes. 13 aggregated probability sample spaces were constructed by varying $T_{1}$ and $T_{2}$. The pairs of $T_{1}$ and $T_{2}$ which were considered are $\{(1,1), (2, \frac{1}{2}), (\frac{1}{2},2),...,(n,\frac{1}{n}),(\frac{1}{n},n),...,(7,\frac{1}{7}),(\frac{1}{7},7)\}$. By varying $T_3$ and $T_4$ and choosing the best pair of expected stopping time and probability of error, the graphs Algo-3 and Algo-4 were obtained in Figure \ref{fig:Figure8}. The construction of the aggregated probability space (\ref{Alternative Decentralized Approach Probability Space Construction}) is helpful, since for a given expected stopping time the probability of error achieved by the second decentralized scheme(\ref{Alternative Decentralized Approach Decision Scheme}) is lower than the probability of error achieved by the first  decentralized scheme (\ref{Solution Decentralized Approach}). As discussed in section \ref{Alternative Decentralized Approach with greater than 1 Bit Exchange},  the performance of the decentralized scheme with greater than 1 bit exchange (Figure \ref{fig:Figure8}, Algo-4) is similar to that of the  decentralized scheme with the construction of the aggregated probability space (Figure \ref{fig:Figure8}, Algo-3), since observations received by the observers are independent conditioned on the hypothesis. Thus there is a trade off between the following: (i) repeated exchange of observations for finding the joint distribution and better performance (than decentralized schemes) in the hypothesis testing problem; (ii) exchange of real valued information only during hypothesis testing and lower performance (than centralized scheme) in the hypothesis testing problem.

Consider the scenario where both observers know the joint distribution of the observations. When Observer 1 needs to compute $\alpha^{1}_{n}$, it needs to find the conditional probability of receiving $Y_{n} = y_{n}$ and $D^{2}_{n}= d^{2}_{n}$ given its own past observations $Y_{1},...,Y_{n-1}$ and the past decisions it receives from Observer 2 $D^{2}_{1},...,D^{2}_{n-1}$.  This computation can be carried out in more than two ways. The first approach would be to search over the observation space, $Y^{n} \times Z^{n}$ for sequences which lead to observed observation and decision pairs ($(Y_{1}, D^{2}_{1}),...,(Y_{n}, D^{2}_{n})$) and then use the joint distribution with the appropriate sequences to find the conditional probability. This is not an efficient approach as computation time increases exponentially with increase in the number of samples. An alternate approach would be to store the sequences found at stage $n$ and then use them to find the sequences at stage $n+1$. In this approach the memory used for storage increases exponentially. Hence even upon knowing the joint distribution of the observations, the  computation of $\alpha^{1}_{n}$ is intensive. For the fourth approach, Observer $i$ needs to  compute $\beta^{i}_{n}$ which requires the joint distribution of $D^{i}_{1},..., D^{i}_{n}$,  and $H$. Again, each observer needs to search over its observation space for finding the observation sequences which lead to that particular decision sequence. Since this approach is computationally intensive, the joint distribution of the decisions was estimated by the frequentist approach. For each observer, $2 \times 2^7 = 256$  decision sequences are possible. From $2 \times 10^7$ samples, the joint distribution of the decision sequence and hypothesis is estimated. 

\begin{figure}
\includegraphics[width=8.4cm,height=5cm]{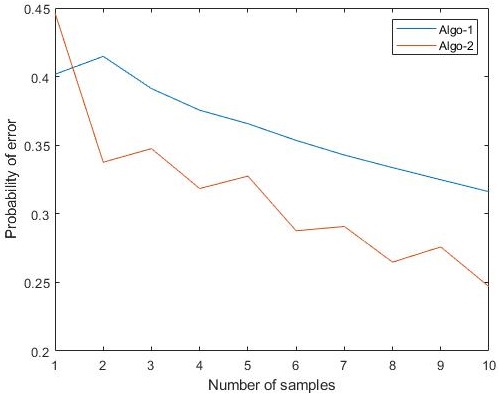}
\caption{Probability of error / conditional probability of agreement on wrong belief )  vs number of samples} 
\label{fig:Figure9}
\includegraphics[width=8.4cm,height=5cm]{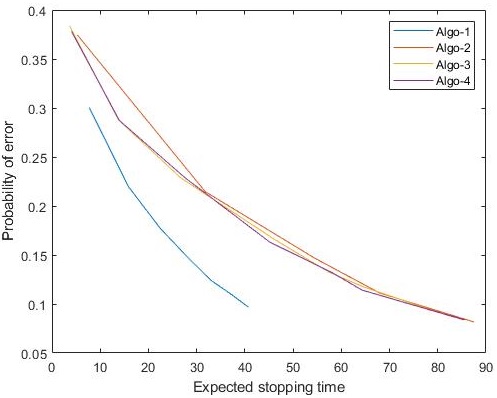}
\caption{Probability of error vs expected stopping time} 
\label{fig:Figure10}
\end{figure}

We considered another setup, where the cardinality of the sets of observations collected by Observers 1 and 2 are $2$ and $3$ respectively. The joint distribution of the observations under either hypothesis is given in Table 2. Under either hypothesis, the observations received by the two observers are not independent. 
\begin{table}[h!]
\begin{center}
\begin{tabular}{||c || c|| c ||c||} 
 \hline
 $f_{0}(y,z)$  & $Z=1$ & $Z=2$ &$Z =3$\\
 \hline
 $Y=1$ & $0.1$ &$0.15$ & $0.2$ \\
 \hline
 $Y=2$ & $0.15$ &$0.2$ & $0.2$ \\
 \hline
\end{tabular}
\begin{tabular}{||c || c|| c ||c ||} 
 \hline
 $f_{1}(y,z)$  & $Z=1$ & $Z=2$ &$Z =3$\\
 \hline
 $Y=1$ & $0.15$ &$0.15$ & $0.25$ \\
 \hline
 $Y=2$ & $0.18$ &$0.14$& $0.13$\\
 \hline
\end{tabular}
\caption{Table 2. Joint distribution of observations under either hypothesis }
\label{table:2}
\end{center}
\vspace{-1cm}
\end{table} 
The prior distribution of the hypothesis was considered to be $p_{0} =0.4$ and $p_{1} =0.6$. $\mathbb{D}_{KL}(f_{1}||f_{0}) = 0.0627 $ and $\mathbb{D}_{KL}(f_{0}||f_{1}) = 0.0649 $. The empirical probability of error achieved by using the centralized scheme as $n$ increases has been plotted in Figure \ref{fig:Figure9} (Algo-1). The empirical probability of the observers agreeing on the wrong belief conditioned on the observers agreeing in the decentralized scheme has been plotted in Figure \ref{fig:Figure9} (Algo-2). $ 2 \times 10^7$ samples were used to construct the aggregated probability space, while the maximum number of possible sequences is $2 \times 2^7 \times 3^7 =559872$. The empirical probability of error achieved by the  centralized sequential hypothesis testing scheme (using the sequential probability ratio test), by the decentralized scheme in section \ref{Solution Decentralized Approach}, by the decentralized scheme in section \ref{Alternative Decentralized Approach Decision Scheme}, by the decentralized scheme in section \ref{Alternative Decentralized Approach with greater than 1 Bit Exchange}, have been plotted against the expected stopping time in Figure \ref{fig:Figure10}, Algo-1, Algo-2, Algo-3, and Algo-4 respectively. There is a significant difference between performance of the centralized and the decentralized schemes. One possible reason is that the marginal distributions are closer, i.e.,  $\mathbb{D}_{KL}(f^{1}_{1}||f^{1}_{0})=0.0290$ and $\mathbb{D}_{KL}(f^{2}_{0}||f^{2}_{1})=0.0244$. The performance of the first decentralized scheme ( \ref{Solution Decentralized Approach}) and the second decentralized scheme are almost similar. Hence the construction of the aggregated probability space is not helpful in this example. 
\section{Conclusion and Future work}\label{Conclusion and Future work}
To conclude, we considered the problem of collaborative binary hypothesis testing with no prior joint distributions knowledge available to the observers. We presented different algorithms to solve the problem with emphasis on information exchange between the observers. The algorithms were analyzed and compared from different perspectives. We proved that the probability space constructed at the observers were a function of information patterns at the observers.  We compared the performance of the algorithms by comparing the rate of decay of the probability of error and proved it is a function of the information exchanged. 

The binary hypothesis testing problem with two observers and asymmetric information can also be studied as a co-operative game with two observers. In future work, we plan to develop game theoretic approaches to this problem following the methods of Grunwald and Topsoe in  \cite{grunwald2004game}, \cite{topsoe2001basic}, \cite{harremoes2002unified}, and \cite{topsoe2009game}.
\bibliographystyle{IEEEtran}
\bibliography{Biblio} 
\section{Appendix}\label{Appendix}
\subsection{Kolmogorov Consistency Theorem}\label{Kolmogorov Consistency Theorem}
This theorem has been invoked in subsections, \ref{Algorithm-1 Learning Phase}, \ref{Algorithm-2 Learning Phase}, \ref{Alternative Decentralized Approach Probability Space Construction}, \ref{Alternative Decentralized Approach Discussion}, and has been mentioned in subsection \ref{Motivation}. We mention the statement of the theorem for the completeness of the paper and for proof we refer to \cite{koralov2007theory}, \cite{billingsley2017probability}. Let $E$ be a parameter set (usually $E = \mathbb{N}, \mathbb{R}_+$) and $(\mathcal{X}, \mathcal{T}_{\mathcal{X}})$ be a topological space (usually $\mathcal{X} = \mathbb{R}, \mathbb{R}^n$ or a finite set). Let $\tilde{\Omega} =\{\omega| \omega: E \to \mathcal{X} \}$. Let $\mathcal{B}(\mathcal{X}^n)$ be a $\sigma$-algebra of subsets of $\mathcal{X}^n$. Let $\{t_1, \ldots, t_n\} \subset E$ and $B \in \mathscr{B}(\mathcal{X}^n)$. A cylindrical subset of $\tilde{\Omega}$ is 
\begin{align*}
    I_{\{t_k\}^n_{k=1}}(B) = \big\{\omega \in  \tilde{\Omega}: \big(\omega(t_1), \ldots, \omega(t_n) \big) \in B   \big\}.
\end{align*}
The collection of all cylindrical subsets for which $\{t_k\}^n_{k=1}$ is fixed and $B$ is allowed to vary is a $\sigma$-algebra denoted by $\mathcal{F}_{\{t_k\}^n_{k=1}}$. Let $\mathcal{B}( \tilde{\Omega})$ be the smallest $\sigma$-algebra containing $\mathcal{F}_{\{t_k\}^n_{k=1}}$, for all $n, \{t_k\}^n_{k=1}$. A collection of probability measures, $\mathbb{P}_{\{t_k\}^n_{k=1}}$ on $\mathcal{F}_{\{t_k\}^n_{k=1}}$, satisfy the consistency conditions if
\begin{itemize}
    \item For every permutation, $\pi$, for every $\{t_1, \ldots, t_n\}$, and $B \in  \mathscr{B}(\mathcal{X}^n)$, 
    \begin{align*}
        \mathbb{P}_{\{t_k\}^n_{k=1}}(I_{\{t_k\}^n_{k=1}}(B)) =  \mathbb{P}_{\pi(\{t_k\}^n_{k=1})}(I_{\{t_k\}^n_{k=1}}(B)) 
    \end{align*}
    \item For every $\{t_1, \ldots, t_n, t_{n+1}\} \subset E$, $B \in  \mathscr{B}(\mathcal{X}^n)$, 
    \begin{align*}
         \mathbb{P}_{\{t_k\}^n_{k=1}}(I_{\{t_k\}^n_{k=1}}(B)) =  \mathbb{P}_{\{t_k\}^{n+1}_{k=1}}(I_{\{t_k\}^{n+1}_{k=1}}(B \times \mathcal{X}))
    \end{align*}
\end{itemize}
The \textit{Kolmogorov Consistency Theorem} or \textit{Kolmogorov Existence Theorem} is stated as, 
\begin{theorem}
    Assume that there exists a family of probability measures, $\{\mathbb{P}_{\{t_k\}^n_{k=1}}\}$,  which satisfy the  consistency conditions. Then, there exists a unique $\sigma$-addtive measure $\tilde{\mathbb{P}}$ on $\mathcal{B}( \tilde{\Omega})$ whose restriction to any $\mathcal{F}_{\{t_k\}^n_{k=1}}$ coincides with $\mathbb{P}_{\{t_k\}^n_{k=1}}$.
\end{theorem}
In this paper, $E = \mathbb{N}$ and $\mathcal{X}$ is a finite set. The distributions found by the observers are for $E_n = \{1,\ldots,n\}, n \in \mathbb{N}$. Hence the distributions are available for only one permutation of the time stamps, specifically, $t_1 < t_2 < \ldots < t_n$. Other permutations are not found as they are not needed for the problem considered. Hence, the first condition of consistency is skipped when the theorem is invoked. 
\subsection{Method of Types}
In this subsection, we recall one result from the method of types \cite{cover2012elements}, \cite{csiszar1998method}, which will be used in subsequent proofs. Let $(Y^{n},Z^{n}) = [(Y_{1},Z_{1}),..., (Y_{n},Z_{n})]$. For an observation sequence $(Y^{n},Z^{n}= y^n,z^n)$, the type associated with it is :
\begin{align*}
\mathbb{Q}_{Y^{n},Z^{n}}(y,z) = \frac{1}{n}\sum^{n}_{i=1}\chi_{ (y,z)}(y_{i},z_{i}), \; \forall (y,z) \in S_{1} \times S_{2}.
\end{align*} 
With the above definition, when $(Y_{1},Z_{1}),...,(Y_{n},Z_{n})$ are i.i.d. conditioned on the hypothesis, for $h = 0,1$,  
\begin{align*}
\mathbb{P}_{n}(Y^{n},Z^{n} = y^n,z^n | H = h) = 2^{-n(\mathbb{H}(\mathbb{Q}_{Y^{n},Z^{n}})+ \mathbb{D}_{KL}(\mathbb{Q}_{Y^{n},Z^{n}} ||f_{h}))}.
\end{align*}
Let $T_{U} = \underset{(y,z) \in S_{1} \times \S_{2}} \max \log_{2} \frac{f_{1}(y,z)}{f_{0}(y,z)}$ and $T_{L} = \underset{(y,z) \in S_{1} \times \S_{2}} \min \log_{2} \frac{f_{1}(y,z)}{f_{0}(y,z)}$. For threshold $T$ such that $T_{L} < log_{2}T < T_{U}$ the likelihood ratio test  can be equivalently written as :
\begin{align*}
\mathbb{D}_{KL}(\mathbb{Q}_{Y^{n},Z^{n}} ||f_{0}) - \mathbb{D}_{KL}(\mathbb{Q}_{Y^{n},Z^{n}} ||f_{1}) \geq \frac{1}{n} \log_{2} T .
\end{align*}
\subsection{Proof of Equations (\ref{Equation 4}) and (\ref{Equation 5})}\label{Appendix Centralized  Approach}
We present the proof for Equations (\ref{Equation 4}) and (\ref{Equation 5}) in subsection \ref{Comparison of Error Rates Centralized Approach}.  
\begin{proof}
Let $\mathcal{S}$ denote the set of probability distributions on $S_{1} \times S_{2}$. For a vector $Q \in \mathcal{S}$, $Q = [Q(1), Q(2), \ldots, Q(|S_{1}| \times |S_{2}|)]$, the element $Q(i)$ corresponds to the joint probability of observing $y_{l}$ and $z_{k}$, where $ l =\lceil \frac{i}{|S_{2}|}\rceil$,  $k = i - \lfloor\frac{i}{|S_{2}|}\rfloor \times |S_{2}|$.  If $i - \lfloor\frac{i}{|S_{2}|}\rfloor \times |S_{2}| = 0$, then $k =|S_{2}|$. $Q(i)$ and $Q(y,z)$ are used interchangeably. For a set $S$, let $int(S)$ denote the interior of the set and $\overline{S}$ denote the closure set. 
Let, 
\begin{align*}
V =	 \left[ \log_{2}\frac{f_{1}(y_{1}, z_{1})}{f_{0}(y_{1},z_{1})},\log_{2}\frac{f_{1}(y_{1}, z_{2})}{f_{0}(y_{1},z_{2})}, \ldots, \log_{2}\frac{f_{1}(y_{|S_{1}|}, z_{|S_{2}|})}{f_{0}(y_{|S_{1}|},z_{|S_{2}|})}\right].
\end{align*} 
For the given threshold $T$, the objective is to find the rate of decay of the probability of error. The set of distributions for which the decision in the centralized case is $1$ is 
\[
 \mathbb{S}_{1}  = \mathbb{Q} \in \mathcal{S} \ni   \left\{ \begin{array}{l} \hspace{-0.2cm}
        \mathbb{D}_{KL}(\mathbb{Q} ||f_{0}) - \mathbb{D}_{KL}(\mathbb{Q} ||f_{1}) \geq \log_{2} T \hspace{-0.2cm}
        \end{array} \right\},
 \]
Let $e_{i}(e_{y,z}), 1 \leq i \leq |S_{1}| \times |S_{2}| $ represent the canonical basis of $\mathbb{R}^{|S_{1}| \times |S_{2}|}$. The set $\mathbb{S}_{1} $ can also be described as: 
\begin{align*}
\mathbb{S}_{1} = \{ &Q \in \mathbb{R}^{|S_{1}| \times |S_{2}|} :  -V^{T}Q + \log_{2}T \leq 0,\\
& \sum_{y,z}Q(y,z) = 1, -e_{i}Q \leq 0, \; 1 \leq i \leq  |S_{1}| \times |S_{2}|\}
\end{align*}
Since $T_{L} < log_{2}T  < T_{U} $, $int(\mathbb{S}_{1}) \neq \emptyset$ and $int(\mathbb{S}^{c}_{1}) \neq \emptyset$. Since $\mathbb{S}_{1}$ and $\overline{\mathbb{S}^{c}_{1}}$ are closed, connected sets with nonempty interiors they are regular closed sets i.e., $\mathbb{S}_{1} = \overline{int(\mathbb{S}_{1})}$ and $\overline{\mathbb{S}^{c}_{1}} = \overline{int(\overline{\mathbb{S}^{c}_{1}})}$. Thus, by By \textit{Sanov's} theorem \cite{cover2012elements}, it follows that
\begin{align*}
&\underset{ n \rightarrow \infty} \lim -\frac{1}{n} \log_{2} (\kappa_{n}) = \mathbb{D}_{KL}(\mathbb{Q}^{0}_{\tau_{0}} || f_{0}),\\
&\underset{ n \rightarrow \infty} \lim -\frac{1}{n} \log_{2} (\xi_{n}) = \mathbb{D}_{KL}(\mathbb{Q}^{1}_{\tau_{1}} || f_{1}),\\
&\mathbb{Q}^{0}_{\tau_{0}}  = \underset{Q \in \mathbb{S}_{1}} \argmin \; \mathbb{D}_{KL}(Q || f_{0}),\mathbb{Q}^{1}_{\tau_{1}}  = \underset{Q \in \overline{\mathbb{S}^c_{1}}} \argmin \; \mathbb{D}_{KL}(Q || f_{0}).
\end{align*}
Since the optimization problems are convex, to solve them the Lagrangian can be setup as follows: 
\begin{align*}
&\mathbb{K}_{h}(Q(y,z),\tau_{h}, \upsilon_{h},\varepsilon_{h}) = \left[\sum_{y,z}Q(y,z)\log_{2}\left(\frac{Q(y,z)}{f_{h}(y,z)}\right) \right] + \\
&s(h)\tau_{h}\left[\sum_{y,z}Q(y,z) \log_{2}\left(\frac{f_{1}(y,z)}{f_{0}(y,z)}\right) -  \log_{2} T\right] - \\ 
&\left[\sum_{y,z}\upsilon_{h}(y,z)e^{T}_{y,z}Q(y,z)\right] + \varepsilon_{h}\left[\sum_{y,z}Q(y,z) - 1 \right]. 
\end{align*}
where $s(h) = -1$ if $h =0$ and $s(h) = 1$ if $h =1$. Setting $\frac{\partial \mathbb{K}_{h}(Q,\tau_{h}, \upsilon_{h},\varepsilon_{h} )}{\partial Q(y,z)} = 0$, for $(y,z) \in S_{1} \times S_{2}$, we get
\begin{align*}
&\log_{2} \left(\frac{Q(y,z)}{f_{h}(y,z)}\right) -s_{h} \tau_{h} \log_{2}\left(\frac{f_{1}(y,z)}{f_{0}(y,z)}\right) + \varepsilon_{h} - \upsilon_{h}(y,z)= -1. \\
& \log_{2} \left( \frac{Q(y,z)\left(f_{0}(y,z)\right)^{s(h)\tau_{h}}}{f_{h}(y,z)\left(f_{1}(y,z)\right)^{s(h)\tau_{h}}}\right)  = - \varepsilon_{h} - 1 +\upsilon_{h}(y,z).
\end{align*}  
Hence, Equation (\ref{Equation 4}) in subsection \ref{Comparison of Error Rates Centralized Approach} follows. The dual functions for the above optimization problems are:
\begin{align*}
\mathbb{J}_{h}(\tau_{h},\upsilon_{h},\varepsilon_{h}) = \mathbb{K}_{h}(\mathbb{Q}^{h}_{\tau_{h}},\tau_{h}, \upsilon_{h},\varepsilon_{h}),
\end{align*}
and the dual optimization problems are:
\begin{align*}
&\Delta^{*}_{h} = \underset{\tau_{h} \in \mathbb{R}, \upsilon_{h} \in \mathbb{R}^{|S{1}| \times |S_{2}|}, \varepsilon_{h} \in \mathbb{R} }  \max \mathbb{J}_{h}(\tau_{h},\upsilon_{h},\varepsilon_{h})\\
& s.t \;\;\; -\tau_{h} \leq 0, -e_{i}\upsilon_{h} \leq 0, \; 1 \leq i \leq  |S_{1}| \times |S_{2}|
\end{align*}
Since the interior of the sets $\mathbb{S}_{1}$ and $\mathbb{S}^{c}_{1}$ are non empty, \textit{Slater's} condition holds and hence strong duality holds. Suppose $\tau^{*}_{h}$ is such that: 
\begin{align}
&\frac{d}{d\tau_{h}}\left[\sum_{y,z}\mathbb{Q}^{h}_{\tau_{h}}(y,z)\log_{2}\left(\frac{\mathbb{Q}^{h}_{\tau_{h}}(y,z)}{f_{h}(y,z)}\right) + s(h)\tau_{h}\times \right. \nonumber \\
& \left.   \left[\sum_{y,z}\mathbb{Q}^{h}_{\tau_{h}}(y,z)\log_{2}\left(\frac{f_{1}(y,z)}{f_{0}(y,z)}\right) \right] \right]\Bigg \vert_{\tau_{h} = \tau^{*}_{h}}  = s(h)\log_{2}T. \label{Equation 11}
\end{align}
Then, since strong duality holds, 
\begin{align*}
\underset{ n \rightarrow \infty} \lim -\frac{1}{n} \log_{2} &(\kappa_{n}) = \Delta^{*}_{0},\underset{ n \rightarrow \infty} \lim -\frac{1}{n} \log_{2} (\xi_{n}) = \Delta^{*}_{1}, \\
&\Delta^{*}_{h} = \mathbb{J}_{h}(\tau^{*}_{h},0, 0)
\end{align*}
Thus, for the given threshold $T$, the rate of decay of probability of error is:
\begin{align*}
\underset{ n \rightarrow \infty} \lim -\frac{1}{n} \log_{2} (\gamma_{n}) = \min \left[\mathbb{D}_{KL}(\mathbb{Q}^{0}_{\tau^{*}_{0}} || f_{0}), \mathbb{D}_{KL}(\mathbb{Q}^{1}_{\tau^{*}_{1}} || f_{1})\right].
\end{align*}
By changing the threshold T (or equivalently $\tau_{0}$ and $\tau_{1}$) different decay rates can be achieved. Thus the optimal rate of decay is achieved by searching over pairs $(\tau_{0}, \tau_{1})$ such that $\tau_{0} \geq 0$ and $\tau_{1} \geq 0$. Further if $R^{*}_{c}$ is achieved by the pair $\bar{\tau}_{0}, \bar{\tau}_{1}$,i.e., 
\begin{align*}
R^{*}_{c} = \min \left[\mathbb{D}_{KL}(\mathbb{Q}^{0}_{\bar{\tau}_{0}}, || f_{0}), \mathbb{D}_{KL}(\mathbb{Q}^{1}_{\bar{\tau}_{1}} || f_{1})\right],
\end{align*}
then $R^{*}_{c} = \mathbb{D}_{KL}(\mathbb{Q}^{0}_{\bar{\tau}_{0}}, || f_{0})$ or $R^{*}_{c} = \mathbb{D}_{KL}(\mathbb{Q}^{1}_{\bar{\tau}_{1}} || f_{1})$. The threshold which achieves the optimal decay rate is found by evaluating the L.H.S of Equation (\ref{Equation 11}) at the appropriate $\bar{\tau}_{h}$ (i.e., the one that achieves $R^{*}_{c}$).  
\end{proof}
\subsection{Proof of Equations (\ref{Equation 8}) and (\ref{Equation 9})}\label{Appendix Decentralized Approach}
In the decentralized scenario, the observation sequence $(Y^{n},Z^{n}= y^n,z^n)$ induces a type on $S_{1}$ and $S_{2}$:
\begin{align*}
\mathbb{Q}^{1}_{Y^{n}}(y) = \frac{1}{n}\sum^{n}_{i=1}\chi_{y}(y_{i}) = \sum_{z \in S_{2}} \mathbb{Q}_{Y^{n},Z^{n}}(y,z), \;\forall\; y \in S_{1}, \\
\mathbb{Q}^{2}_{Z^{n}}(z) = \frac{1}{n}\sum^{n}_{i=1}\chi_{z}(z_{i}) = \sum_{y \in S_{1}} \mathbb{Q}_{Y^{n},Z^{n}}(y,z), \;\forall\; z \in S_{2}.
\end{align*}
Let, 
\begin{align*}
&T^{1}_{U} = \underset{ y \in S_{1}} \max \log_{2} \frac{f^{1}_{1}(y)}{f^{1}_{0}(y)}, \; \; T^{2}_{U}=\underset{ z \in S_{2}} \max \log_{2} \frac{f^{2}_{1}(z)}{f^{2}_{0}(z)} \\
&T^{1}_{L} = \underset{ y \in S_{1}} \min \log_{2} \frac{f^{1}_{1}(y)}{f^{1}_{0}(y)}, \; \;T^{2}_{L} = \underset{ z \in S_{2}} \min \log_{2} \frac{f^{2}_{1}(z)}{f^{2}_{0}(z)}
\end{align*}
Let $T_{1}$ and $T_{2}$ be such that $T^{1}_{L} < \log_{2} T_{1} < T^{1}_{U}$ and $T^{2}_{L}< \log_{2}T_{2} < T^{2}_{U}$. The individual likelihood ratio tests for the observers with thresholds $T_{1}$ and $T_{2}$ are :
\begin{align*}
\mathbb{D}_{KL}(\mathbb{Q}^{1}_{Y^{n}} ||f^{1}_{0}) - \mathbb{D}_{KL}(\mathbb{Q}^{1}_{Y^{n}} ||f^{1}_{1}) \geq \frac{1}{n} \log_{2} T_{1},\\
\mathbb{D}_{KL}(\mathbb{Q}^{2}_{Z^{n}} ||f^{2}_{0}) - \mathbb{D}_{KL}(\mathbb{Q}^{2}_{Z^{n}} ||f^{2}_{1}) \geq \frac{1}{n} \log_{2} T_{2}.
\end{align*}
Now, we present the proof for Equation (\ref{Equation 9}) in subsection \ref{Comparison of Error Rates Decentralized Approach}.
\begin{proof} 
Let,
\begin{align*}
&v = [1,1,..., 1] \in \mathbb{R}^{|S_{2}|},\; v_{1} =[1,1, \ldots, 1] \in \mathbb{R}^{|S_{1}| \times |S_{2}|}\\
&u = \left[\log_{2}\frac{f^{2}_{1}(z_{1})}{f^{2}_{0}(z_{1})}, \log_{2}\frac{f^{2}_{1}(z_{2})}{f^{2}_{0}(z_{2})}, ..., \log_{2}\frac{f^{2}_{1}(z_{|S_{2}|})}{f^{2}_{0}(z_{|S_{2}|})}\right] \in \mathbb{R}^{|S_{2}|}, \\
&v_{2} = \left[\log_{2}\frac{f^{1}_{1}(y_{1})}{f^{1}_{0}(y_{1})} \times v,\log_{2}\frac{f^{1}_{1}(y_{2})}{f^{1}_{0}(y_{2})} \times v,...,\log_{2}\frac{f^{1}_{1}(y_{|S_{1}|})}{f^{1}_{0}(y_{|S_{1}|})} \times v \right]\\
& \in \mathbb{R}^{|S_{1}| \times |S_{2}|}, \; v_{3}=[u,u,...,u] \in \mathbb{R}^{|S_{1}| \times |S_{2}|},||Q||_{\infty} = \\ 
&\underset{i} \max \; |Q(i)| ,
 Q \in \mathbb{R}^{|S_{1}| \times |S_{2}|}, \; M_{1} = \left[\sum_{y \in S_{1}} \left|\log_{2} \frac{f^{1}_{1}(y)}{f^{1}_{0}(y)}\right | \right]\times |S_{2}|.
\end{align*}
For the given pair of thresholds $T_{1}, T_{2}$,  the objective is to find the rate of decay of the probability of false alarm and of the probability of miss detection. We first focus on the rate of decay of the probability of false alarm. The set of distributions for which the decisions of both observers is $1$ is
\[
 \mathcal{S}_{1}  = \mathbb{Q} \in \mathcal{S} \ni   \left\{ \begin{array}{l} \hspace{-0.2cm}
        \mathbb{D}_{KL}(\mathbb{Q}_{1} ||f^{1}_{0}) - \mathbb{D}_{KL}(\mathbb{Q}_{1} ||f^{1}_{1}) \geq \log_{2} T_{1} \hspace{-0.2cm} \\
 \hspace{-0.2cm} \mathbb{D}_{KL}(\mathbb{Q}_{2} ||f^{2}_{0}) - \mathbb{D}_{KL}(\mathbb{Q}_{2}||f^{2}_{1}) \geq  \log_{2} T_{2} \hspace{-0.2cm} 
        \end{array} \right\},
 \]
where $\mathbb{Q}_{1}$ and $\mathbb{Q}_{2}$ are types induced by $\mathbb{Q}$ on $S_{1}$ and $S_{2}$ respectively.  The set $\mathcal{S}_{1} $ can also be described as :
\begin{align*}
\mathcal{S}_{1} = \{ &Q \in \mathbb{R}^{|S_{1}| \times |S_{2}|} :  -v_{2}^{T}Q + \log_{2}T_{1} \leq 0,  v_{1}^{T}Q = 1,  \\
&  -v_{3}^{T}Q + \log_{2}T_{2} \leq 0, \; \; -e_{i}Q \leq 0, \; 1 \leq i \leq  |S_{1}| \times |S_{2}|\}
\end{align*}
The first objective is to find threshold pairs $T_{1}, T_{2}$ for which $\mathcal{S}_{1}$ is non empty. Note that,
\begin{align*}
&\underset{Q \in \mathcal{S}} \max v_{2}^{T}Q  = \underset{ y \in S_{1}} \max \log \frac{f^{1}_{1}(y)}{f^{1}_{0}(y)}  \; , \; \underset{Q \in \mathcal{S}} \max v_{3}^{T}Q = \underset{ z \in S_{2}} \max \log \frac{f^{2}_{1}(z)}{f^{2}_{0}(z)}, \\
&\underset{Q \in \mathcal{S}} \min v_{2}^{T}Q  = \underset{ y \in S_{1}} \min \log \frac{f^{1}_{1}(y)}{f^{1}_{0}(y)} \; , \; \underset{Q \in \mathcal{S}} \min v_{3}^{T}Q = \underset{ z \in S_{2}} \min \log \frac{f^{2}_{1}(z)}{f^{2}_{0}(z)}. 
\end{align*}
Since $T^{2}_{L} < \log_{2} T_{2} < T^{2}_{U}$, and $g(Q) = v_{3}^{T}Q $ is continuous, $\exists \; Q_{a} \in \mathcal{S} $ such that $ v_{3}^{T}Q_{a}= \log_{2} T_{2}$. For a feasible $T_{2}$, we would like to find the set of feasible $T_{1}$ so that that the set $\mathcal{S}_{1}$ is nonempty. Consider:
\begin{align*}
\Psi(T_{2}) = &\underset{Q \in \mathbb{R}^{|S_{1}| \times |S_{2}|}} \max v_{2}^{T}Q\\
& \text{s.t} \;\; -v_{3}^{T}Q + \log_{2}T_{2} \leq 0, \; v_{1}^{T}Q = 1,\\
&\;\; \;\;\;\;-e_{i}Q \leq 0, \; 1 \leq i \leq  |S_{1}| \times |S_{2}|\\
\Phi(T_{2}) = &\underset{Q \in \mathbb{R}^{|S_{1}| \times |S_{2}|}} \min v_{2}^{T}Q\\
& \text{s.t} \;\; -v_{3}^{T}Q + \log_{2}T_{2} \leq 0, \; v_{1}^{T}Q = 1,\\
&\;\; \;\;\;\;-e_{i}Q \leq 0, \; 1 \leq i \leq  |S_{1}| \times |S_{2}|
\end{align*}
Since the above optimization problems are linear programs for every $T_{2}$, the maximum and the minimum occur at one of the vertices of the convex polygon, $\mathcal{S}_{2} = \mathcal{S} \cap \{ Q:  -v_{3}^{T}Q - \log_{2}T_{2} \leq 0\}$. Let $int(S)$ denote the interior of a set $S$. Let $Q$ be a boundary point of the set $S$. Let $C(Q,S) = \{h: \exists \bar{\epsilon}> 0, \; \text{s.t.} \; Q +\epsilon h \in int(S), \; \forall \epsilon \in [0, \bar{\epsilon}]\} $.

Since the set $\mathcal{S}$ is convex, for any point $Q_{a}$ in the interior of the set and $Q$ on its boundary, the vector $Q_{a}- Q$ belongs to $C(Q, \mathcal{S})$. For a given $T_{1},T_{2}$, if $\Phi(T_{2}) < \log_{2} T_{1} < \Psi(T_{2})$, then the pair is a feasible pair. If not, we choose an alternative $T_{1}$ which satisfies the above inequalities. Further we choose $T$ be such that $\Phi(T_{2}) < \log_{2} T_{1} < \log_{2} T < \Psi(T_{2})$. Since the function $f(Q) = v_{2}^{T}Q$ is continuous, $\exists \; Q_{a} \in \mathcal{S}_{2} $ such that $f(Q_{a}) = \log_{2} T$. Hence $Q_{a} \in \mathcal{S}$ is such that $v_{2}^{T}Q_{a} > \log_{2} T_{1}$ and $v_{3}^{T}Q_{a} \geq \log_{2} T_{2} $, which implies that the set $\mathcal{S}_{1}$ is nonempty. 

If $Q_{a}$ is an interior  point of $\mathcal{S}_{2}$ then it is an interior point for $\mathcal{S}_{1}$. Suppose $Q_{a}$ is a boundary point of $\mathcal{S}_{2}$, such that $v_{3}^{T}Q_{a} = \log_{2}T_{2}$ and $Q_{a}(i) > 0$ for all $i$. There exists a direction $h$ such that $v^{T}_{3}h > 0$ and for epsilon small enough, $(Q_{a} +\epsilon h)$ belongs to the interior of $\mathcal{S}_{2}$. Suppose $Q_{a}$ is a boundary point of $\mathcal{S}_{2}$, such that $Q_{a}(i) = 0$ for some $i$. Then, set $C(Q_{a}, \mathcal{S})\cap \{ h : v^{T}_{3}h \geq 0\}$ is nonempty. 

Indeed, if the set is empty then $C(Q_{a}, \mathcal{S}) \subseteq \{ h : v^{T}_{3} h< 0\}$ which implies that $v^{T}_{3}Q < log_{2}T_{2} \; \forall Q \in int(\mathcal{S})$, which is a contradiction as  $\log_{2} T_{2} < T^{2}_{U}$. This can be proven by the following argument. Let $Q_{c}$ be such that $v^{T}_{3} Q_{c} = T^{2}_{U}$. Note that $Q_{c}$ is a boundary point of $\mathcal{S}$. Let $\epsilon = \frac{T^{2}_{U} - log_{2}T_{2}}{4}$. By continuity of $v^{T}_{3} Q$, there exits $\delta > 0$ such that $||Q- Q_{c}||_{\infty} < \delta $ implies $|v^{T}_{3} Q - v^{T}_{3} Q_{c}| < \epsilon$. This implies for every $Q$ such that $||Q- Q_{c}||_{\infty} < \delta $, $v^{T}_{3} Q > T^{2}_{U} - \epsilon > \log_{2} T_{2}$. Since $Q_{c}$ is a boundary point of $\mathcal{S}$, there exists at least one interior point of $\mathcal{S}$ in the ball, $||Q- Q_{c}||_{\infty} < \delta $. Hence there exists an interior point, $Q_{d}$ such that  $v^{T}_{3} Q_{d} > \log_{2} T_{2}$, which contradicts our conclusion that $v^{T}_{3}Q < log_{2}T_{2} \; \forall Q \in int(\mathcal{S})$. 

\noindent Thus, there exits $Q_{b}$ an interior point of $\mathcal{S}$, such that  $Q_{b}(i) > 0 \; \forall \; i$, $v_{3}^{T}Q_{b} >  \log_{2}T_{2}$, $||Q_{a} - Q_{b}||_{\infty} < \epsilon$ and 
\begin{align*}
v_{2}^{T}Q_{b} &= v^{T}_{2}Q_{a} + v_{2}^{T}Q_{b} - v_{2}^{T}Q_{a}\\
&\geq \log_{2} T - ||Q_{a} - Q_{b}||_{\infty} \times M_{1}\\
&\geq \log_{2} T - \epsilon\times M_{1}.
\end{align*}
We choose $\epsilon$ such that $\epsilon < \frac{\log_{2} T -\log_{2} T_{1}}{2 \times M_{1}}$.   Then, $v_{2}^{T}Q_{b}> \frac{\log_{2} T + \log_{2} T_{1} }{2} > \log_{2} T_{1}$. Hence $Q_{b}$ is an interior point of $\mathcal{S}_{1}$. Thus, for the $T_{1}$, $T_{2}$ pair, there exists $Q \in \mathcal{S}$ such that $Q(i)> 0, \; \forall \; i$, $v^{T}_{2}Q > log_{2}T_{1}$, $v^{T}_{3}Q > log_{2}T_{2}$. Hence the interior of the set $\mathcal{S}_{1}$ is also nonempty. Clearly, $\mathcal{S}_{1}$ is closed and convex. Since $\mathcal{S}_{1}$ is a connected, closed set with nonempty interior, it is a regular closed set ($\mathcal{S}_{1} = \overline{int(\mathcal{S}_{1})}$).[A connected set is a set which cannot be partitioned into two nonempty subsets such that each subset has no points in common with the set closure of the other. Using this definition and a contradiction argument we can show that a closed,connected set with nonempty interior is a regular closed set.]

\noindent By \textit{Sanov's} theorem \cite{cover2012elements}, it follows that :
\begin{align*}
\underset{ n \rightarrow \infty} \lim -\frac{1}{n} \log_{2} (\mu_{n}) = \mathbb{D}_{KL}(\mathbb{Q}^{0}_{\lambda_{0},\sigma_{0}} || f_{0}),
\end{align*}
where,
\begin{align*}
\mathbb{Q}^{0}_{\lambda_{0},\sigma_{0}}  = \underset{Q \in \mathcal{S}_{1}} \argmin \; \mathbb{D}_{KL}(Q || f_{0}).
\end{align*}
To find $\mathbb{Q}^{0}_{\lambda_{0},\sigma_{0}}$, the Lagrangian can be set up as follows :
\begin{align*}
&\mathbb{L}(Q, \lambda_{0},\sigma_{0},\zeta_{0} ,\theta_{0} ) = \left[\sum_{y,z}Q(y,z)\log_{2}\left(\frac{Q(y,z)}{f_{0}(y,z)}\right) \right]+ \\
&\lambda_{0}\left[ \log_{2} T_{1} - \sum_{y}\left(\sum_{z}Q(y,z)\right)\log_{2}\left(\frac{f^{1}_{1}(y)}{f^{1}_{0}(y)}\right) \right] + \\ 
& \sigma_{0}\left[ \log_{2} T_{2} - \sum_{z}\left(\sum_{y}Q(y,z)\right)\log_{2}\left(\frac{f^{2}_{1}(z)}{f^{2}_{0}(z)}\right)\right] -  \\
&\left[\sum_{y,z}\zeta(y,z)e^{T}_{y,z}Q(y,z)\right] + \theta_{0}\left[\sum_{y,z}Q(y,z) - 1 \right]. 
\end{align*}
Setting $\frac{\partial \mathbb{L}(Q, \lambda_{0},\sigma_{0}, \zeta_{0},\theta_{0})}{\partial Q(y,z)} = 0$, for $(y,z) \in S_{1} \times S_{2}$, we get 
\begin{align*}
&\log_{2} \left(\frac{Q(y,z)}{f_{0}(y,z)}\right) - \lambda_{0} \log_{2}\left(\frac{f^{1}_{1}(y)}{f^{1}_{0}(y)}\right) - 
\\ & \sigma_{0}\log_{2}\left(\frac{f^{2}_{1}(z)}{f^{2}_{0}(z)}\right) + \theta_{0} + 1 - \zeta(y,z)= 0. \\
& \log_{2} \left( \frac{Q(y,z)\left(f^{1}_{1}(y)\right)^{-\lambda_{0}}\left(f^{2}_{1}(z)\right)^{-\sigma_{0}}}{f_{0}(y,z)\left(f^{1}_{0}(y)\right)^{-\lambda_{0}}\left(f^{2}_{0}(z)\right)^{-\sigma_{0}}}\right)  = - \theta_{0} - 1 +\zeta(y,z).
\end{align*}  
Hence the definition of $\mathbb{Q}^{0}_{\lambda_{0},\sigma_{0}}$ as in Equation (\ref{Equation 8}), in section \ref{Comparison of Error Rates Decentralized Approach}, follows. The dual function is defined as: 
\begin{align*}
\mathbb{G}(\lambda_{0},\sigma_{0},\zeta_{0},\theta_{0}) = \mathbb{L}(\mathbb{Q}^{0}_{\lambda_{0},\sigma_{0}}, \lambda_{0},\sigma_{0},\zeta_{0} , \theta_{0}). 
\end{align*}
The dual optimization problem is defined as 
\begin{align*}
&d^{*} =  \underset{\lambda_{0} \in \mathbb{R}, \sigma_{0} \in \mathbb{R},\zeta_{0} \in \mathbb{R}^{|S{1}| \times |S_{2}|}, \theta_{0} \in \mathbb{R} }  \max \mathbb{G}(\lambda_{0},\sigma_{0},\zeta_{0},\theta_{0})\\
& \hspace{2.5cm} s.t \;\;\; -\lambda_{0} \leq 0,  -\sigma_{0} \leq 0,\\
& \hspace{2.5cm} -e_{i}\zeta_{0} \leq 0, \; 1 \leq i \leq  |S_{1}| \times |S_{2}|
\end{align*}
Since the interior of the set $\mathcal{S}_{1}$ is nonempty, \textit{Slater's} condition holds and hence strong duality holds. Hence, 
\begin{align*}
\underset{ n \rightarrow \infty} \lim -\frac{1}{n} \log_{2} (\mu_{n}) = d^{*}.
\end{align*}
Suppose $\lambda^{*}_{0}$ and $\sigma^{*}_{0}$ are such that:
\begin{align}
&\frac{\partial }{\partial \lambda_{0}} \left [\left[\sum_{y,z}\mathbb{Q}^{0}_{\lambda_{0},\sigma_{0}}(y,z)\log_{2}\left(\frac{\mathbb{Q}^{0}_{\lambda_{0},\sigma_{0}}(y,z)}{f_{0}(y,z)}\right) \right]- \right. \nonumber \\
&\lambda_{0}\left[\sum_{y}\sum_{z}\mathbb{Q}^{0}_{\lambda_{0},\sigma_{0}}(y,z)\log_{2}\left(\frac{f^{1}_{1}(y)}{f^{1}_{0}(y)}\right) \right] -\nonumber  \\ 
& \left. \sigma_{0}\left[\sum_{z}\sum_{y}\mathbb{Q}^{0}_{\lambda_{0},\sigma_{0}}(y,z)\log_{2}\left(\frac{f^{2}_{1}(z)}{f^{2}_{0}(z)}\right)\right]\right]\Bigg\vert _{\lambda^{*}_{0},\sigma^{*}_{0}} = - \log_{2} T_{1} \nonumber \\
&\frac{\partial }{\partial \sigma_{0}} \left [\left[\sum_{y,z}\mathbb{Q}^{0}_{\lambda_{0},\sigma_{0}}(y,z)\log_{2}\left(\frac{\mathbb{Q}^{0}_{\lambda_{0},\sigma_{0}}(y,z)}{f_{0}(y,z)}\right) \right]- \right. \nonumber \\
&\lambda_{0}\left[\sum_{y}\sum_{z}\mathbb{Q}^{0}_{\lambda_{0},\sigma_{0}}(y,z)\log_{2}\left(\frac{f^{1}_{1}(y)}{f^{1}_{0}(y)}\right) \right] -\nonumber  \\ 
& \left. \sigma_{0}\left[\sum_{z}\sum_{y}\mathbb{Q}^{0}_{\lambda_{0},\sigma_{0}}(y,z)\log_{2}\left(\frac{f^{2}_{1}(z)}{f^{2}_{0}(z)}\right)\right]\right]\Bigg\vert _{\lambda^{*}_{0},\sigma^{*}_{0}} = - \log_{2} T_{2} \label{Equation 12}
\end{align}
By solving the above equations, the optimizers $\lambda^{*}_{0}$ and $\sigma^{*}_{0}$ can be found as functions of $T_{1}$ and $T_{2}$ and the distribution which achieves the optimal rate for this pair of thresholds is $\mathbb{Q}^{0}_{\lambda^{*}_{0},\sigma^{*}_{0}}$. To study the rate of decay of the probability of miss detection we consider the set of distributions for which the decision of both observers is 0, $\mathcal{S}_{3}$, 
\begin{align*}
\mathcal{S}_{3} = \{ &Q \in \mathbb{R}^{|S_{1}| \times |S_{2}|} :  v_{2}^{T}Q - \log_{2}T_{1} \leq 0,  v_{1}^{T}Q = 1,  \\
&  v_{3}^{T}Q - \log_{2}T_{2} \leq 0, \; \; -e_{i}Q \leq 0, \; 1 \leq i \leq  |S_{1}| \times |S_{2}|\}.
\end{align*}
It is clear that $\mathcal{S}_{3}$ is closed, convex and has nonempty interior (as $T^{2}_{L} < T_{2}$ and $\Phi(T_{2}) < \log_{2} T_{1}$). Again by \textit{Sanov's} theorem,
\begin{align*}
\underset{ n \rightarrow \infty} \lim -\frac{1}{n} \log_{2} (\nu_{n}) = \mathbb{D}_{KL}(\mathbb{Q}^{1}_{\lambda_{1},\sigma_{1}} || f_{1}),
\end{align*}
where,
\begin{align*}
\mathbb{Q}^{1}_{\lambda_{1},\sigma_{1}}  = \underset{Q \in \mathcal{S}_{1}} \argmin \; \mathbb{D}_{KL}(Q || f_{1}).
\end{align*}  
The optimization problem can be solved to show that $\mathbb{Q}^{1}_{\lambda_{1},\sigma_{1}}$ satisfies Equation (\ref{Equation 8}) for $h=1$.  The dual problem can be solved to find $\lambda^{*}_{1}$ and $\sigma^{*}_{1}$. Thus for the given thresholds (and hence decision policy), the error rate is 
\begin{align*}
\underset{n \rightarrow \infty} \lim -\frac{1}{n} \log_{2} \left(\rho_{n}\right) =  \min \left[ \mathbb{D}_{KL}(\mathbb{Q}^{0}_{\lambda^{*}_{0},\sigma^{*}_{0}} || f_{0}), \mathbb{D}_{KL}(\mathbb{Q}^{1}_{\lambda^{*}_{1},\sigma^{*}_{1}} || f_{1})\right],
\end{align*}
since the exponential rate is determined by the worst exponent.  By changing the thresholds (and hence $\lambda_{h}$, $\sigma_{h}$, $h=0,1$), different error rates can be obtained. Thus the best error rate is obtained by taking the maximum over $\lambda_{h} \geq 0 $ and $\sigma_{h} \geq 0$, $h=0,1$. Thus, Equation (\ref{Equation 9}), in section \ref{Comparison of Error Rates Decentralized Approach}, follows. Suppose the above maximum is achieved at $(\bar{\lambda}_{0}, \bar{\sigma}_{0}), (\bar{\lambda}_{1}, \bar{\sigma}_{1})$. Then $R^{*}_{d} = \mathbb{D}_{KL}(\mathbb{Q}^{0}_{\bar{\lambda}_{0},\bar{\sigma}_{0}}|| f_{0})$ or $R^{*}_{d} = \mathbb{D}_{KL}(\mathbb{Q}^{1}_{\bar{\lambda}_{1},\bar{\sigma}_{1}},|| f_{1})$. Suppose $R^{*}_{d} = \mathbb{D}_{KL}(\mathbb{Q}^{0}_{\bar{\lambda}_{0},\bar{\sigma}_{0}}|| f_{0})$. Then, the thresholds which achieve the optimal rate of decay can be found by evaluating the L.H.S of Equation (\ref{Equation 12}) at $(\bar{\lambda}_{0}, \bar{\sigma}_{0})$. For the other case, the thresholds can be found from equations analogous to Equation (\ref{Equation 12}), which arise from the dual optimization problem obtained while finding the rate of decay of the probability of miss detection. 
\subsection{Proof of Equation (\ref{Equation 10})}
\begin{figure}
\includegraphics[width=8.4cm,height=5cm]{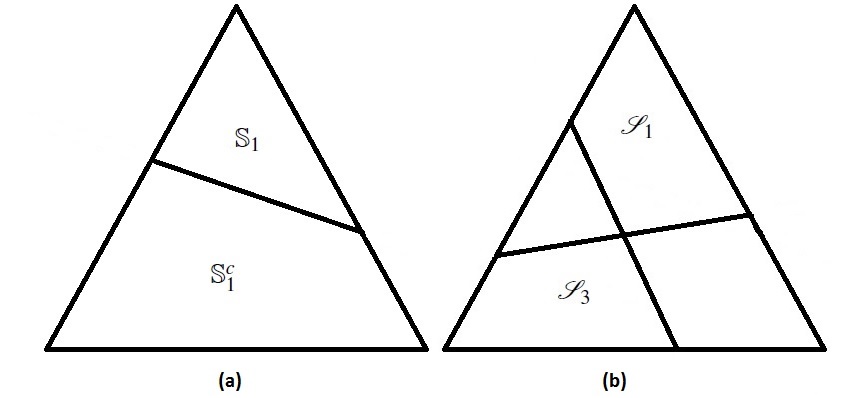}
\caption{Bifurcation of the probability simplex in the two approaches: (a) Centralized (b) Decentralized} 
\label{fig:Figure11}
\vspace{-0.5cm}
\end{figure}
\noindent Suppose the observation collected by Observer 1 is independent of the observation collected by Observer 2 under either hypothesis, i.e., $f_{0}(y,z) = f^{1}_{0}(y)f^{2}_{0}(z)$, $f_{1}(y,z) = f^{1}_{1}(y)f^{2}_{1}(z)$. Let $\mathbb{C}_{1}$ be a subset of the positive cone, $\mathbb{C}_{1} = \{(\lambda_{0}, \sigma_{0}, \lambda_{1}, \sigma_{1} ) \in \mathbb{R}^{4} \lambda_{0}, \sigma_{0}, \lambda_{1}, \sigma_{1} \geq 0,  \lambda_{0} = \sigma_{0}, \lambda_{1} = \sigma_{1}\}$. For such quadruplets, 
\begin{align*}
\mathbb{Q}^{h}_{\lambda_{h},\sigma_{h}} \Big \vert_{\lambda_{h}=\sigma_{h} =\tau_{h}} = \mathbb{Q}^{h}_{\tau_{h}}.
\end{align*}
Thus, 
\begin{align*}
R^{*}_{d} &=  &&\max_{\lambda_{h}\geq 0, \sigma_{h} \geq 0, h= 0, 1} \hspace{-1.2cm}&&&\min \left[\mathbb{D}_{KL}(\mathbb{Q}^{0}_{\lambda_{0},\sigma_{0}} || f_{0}), \mathbb{D}_{KL}(\mathbb{Q}^{1}_{\lambda_{1},\sigma_{1}} || f_{1})\right]\\
& \geq &&\max_{(\lambda_{h}\geq 0, \sigma_{h}, h= 0, 1) \in \mathbb{C}_{1}}  \hspace{-1.2cm} &&&\min \left[\mathbb{D}_{KL}(\mathbb{Q}^{0}_{\lambda_{0},\sigma_{0}} || f_{0}), \mathbb{D}_{KL}(\mathbb{Q}^{1}_{\lambda_{1},\sigma_{1}} || f_{1})\right]\\
&= &&\hspace{0.75cm}\max_{\tau_{0},\tau_{1} \geq 0} \hspace{-1.2cm}&&&\min \left[\mathbb{D}_{KL}(\mathbb{Q}^{0}_{\tau_{0}} || f_{0}), \mathbb{D}_{KL}(\mathbb{Q}^{1}_{\tau_{1}} || f_{1})\right] = R^{*}_{c}
\end{align*}
The above result can be understood as follows: in the centralized case, the probability simplex is divided into two regions by a hyperplane, while in the decentralized case the simplex is divide into four regions by two hyperplanes. Hence, the minimum of the Kullback-Liebler divergence between the decision regions (in the probability simplex) and the observation distributions in the centralized scenario is likely to be lower than in the decentralized case as the sets are ``larger" in the centralized scenario (Figure \ref{fig:Figure11}).
\end{proof}
\vspace{-2.5cm}
\begin{IEEEbiography}[{\includegraphics[width=1.in,height=1.15in,clip,keepaspectratio]{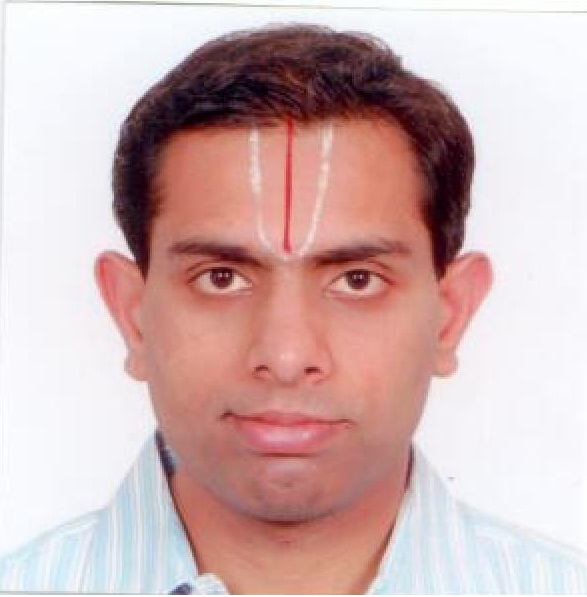}}]{Aneesh Raghavan} received the Bachelor of Technology degree in Instrumentation Engineering from the Indian Institute of Technology, Kharagpur in 2013 and Ph.D. degree in Electrical Engineering from the University of Maryland, College Park, in 2020. Currently, he is a Post-Doc at the Division of Decision and Control Systems, KTH, Royal Institute of Technology, Stockholm. His research interests include collaborative inference, active learning, and, neuro-symbolic reasoning and learning. 
\end{IEEEbiography}
\vspace{-2.5cm}
\begin{IEEEbiography}[{\includegraphics[width=1.1in,height=1.25in,clip,keepaspectratio]{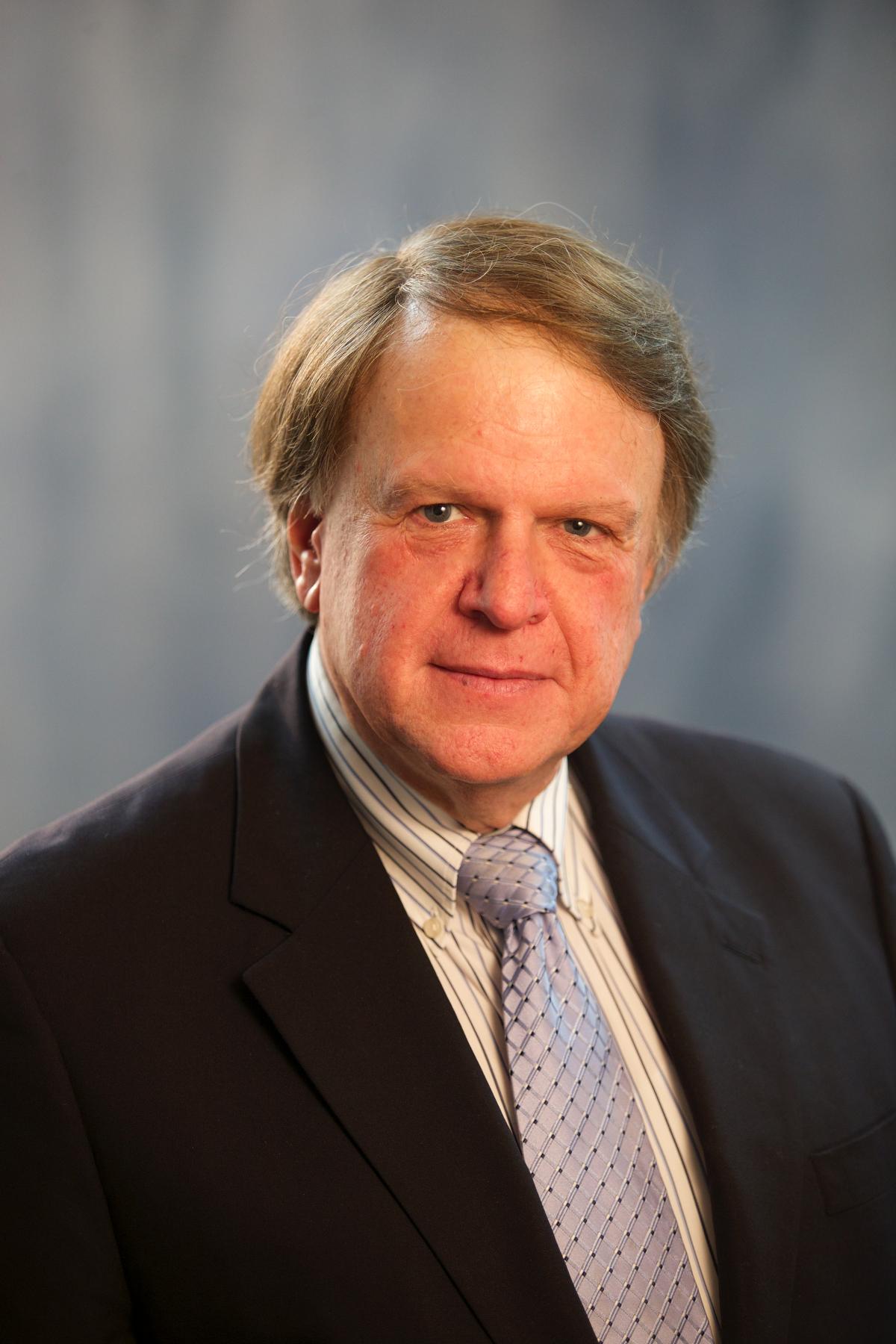}}]{John S. Baras}
received the M.S. and Ph.D. degrees in Applied Mathematics from Harvard University, Cambridge, MA, USA, in 1971 and 1973, respectively. Since 1973, he has been with the Department of Electrical and Computer Engineering, University of Maryland at College Park, MD, USA, where he is currently a Distinguished University Professor, holding a Permanent Joint Appointment with the Institute for Systems Research (ISR) and the Lockheed Martin endowed Chair in Systems Engineering. He is also a Faculty member of the Applied Mathematics, Statistics and Scientific Computation Program, and an Affiliate Professor in the Departments of Computer Science, Fischell Bioengineering, Mechanical Engineering, Aerospace Engineering, Decision, Operations and Information Technologies of the Robert H. Smith School of Business. From 1985 to 1991, he was the Founding Director of the ISR (one of the first six National Science Foundation Engineering Research Centers). Since 1992, he has been the Director of the Maryland Center for Hybrid Networks (HYNET), which he co-founded.

He is an IEEE Life Fellow, and Fellow of SIAM, AAAS, NAI, IFAC, AMS, AIAA, Member of the National Academy of Inventors (NAI) and a Foreign Member of the Royal Swedish Academy of Engineering Sciences (IVA). Major honors and awards include the 1980 George Axelby Award from the IEEE Control Systems Society, the 2006 Leonard Abraham Prize from the IEEE Communications Society, the 2014 Tage Erlander Guest Professorship from the Swedish Research Council, and a three year (2014-2017) Senior Hans Fischer Fellowship from the Institute for Advanced Study of the Technical University of Munich, Germany. In 2016 he was inducted in the University of Maryland A. J. Clark School of Engineering Innovation Hall of Fame. He was awarded the 2017 Institute for Electrical and Electronics Engineers (IEEE) Simon Ramo Medal, the 2017 American Automatic Control Council (AACC) Richard E. Bellman Control Heritage Award, and the 2018 American Institute for Aeronautics and Astronautics (AIAA) Aerospace Communications Award. In 2018 he was awarded a Doctorate Honoris Causa by the National Technical University of Athens, Greece. 

He has educated 102 doctoral students, 160 MS students and has mentored 70 postdoctoral fellows. He has given many plenary and keynote addresses in major international conferences worldwide. He has been awarded nineteen patents and has been honored worldwide with many awards as innovator and leader of economic development. 

\end{IEEEbiography}

\end{document}